\documentclass[a4paper,fleqn,usenatbib]{mnras}

\usepackage{newtxtext,newtxmath}
\usepackage[T1]{fontenc}
\usepackage{ae,aecompl}

\usepackage{graphicx}	% Including figure files
\usepackage{amsmath}	% Advanced maths commands
\usepackage{amssymb}	% Extra maths symbols
\usepackage{array}
\usepackage{graphicx,nicefrac,url}
\usepackage{wasysym}

\newcolumntype{C}[1]{>{\centering\let\newline\\\arraybackslash\hspace{0pt}}m{#1}}

\newcommand{\bes}{Besan\c{c}on}
\newcommand{\mabuls}{MaB$\mu$lS}
\newcommand{\kms}{km~s$^{-1}$}
\newcommand{\msyr}{mas~year$^{-1}$}

\graphicspath{ {figures/} }

\title{\mabuls{}-2: high-precision microlensing modelling for the large-scale survey era}

\author[D. Specht et al.]{
David Specht,$^{1}$\thanks{E-mail: david.specht@manchester.ac.uk}
Eamonn Kerins,$^{1}$\thanks{E-mail: eamonn.kerins@manchester.ac.uk}
Supachai Awiphan$^{2}$\thanks{E-mail: supachai@narit.or.th}
and Annie C. Robin$^{3}$\thanks{E-mail: annie.robin@obs-besancon.fr}
\\
$^{1}$Jodrell Bank Centre for Astrophysics, The University of Manchester, M13 9PL, Manchester, UK\\
$^{2}$National Astronomical Research Institute of Thailand, 260 Moo 4, T. Donkaew,  A. Maerim, Chiangmai, 50180 Thailand\\
$^{3}$Institut Utinam, CNRS UMR 6213, Univ. Bourgogne Franche-Comté, OSU THETA, Observatoire de \bes{}, BP 1615 25010 \bes{} Cedex, France
}

\date{Accepted 2020 August 6. Received 2020 July 28; in original form 2020 May 28}

\pubyear{2020}

\begin{document}
\label{firstpage}
\pagerange{\pageref{firstpage}--\pageref{lastpage}}
\maketitle

\begin{abstract}
Galactic microlensing datasets now comprise in excess of $10^4$ events, and with the advent of next generation microlensing surveys that may be undertaken with facilities such as the Rubin Observatory (formerly LSST) and Roman Space Telescope (formerly WFIRST), this number will increase significantly. So too will the fraction of events with measurable higher order information such as  finite source effects and lens--source relative proper motion. Analysing such data requires a more sophisticated Galactic microlens modeling approach. We present a new second-generation Manchester--\bes{} Microlensing Simulator (\mabuls{}-2), which uses a version of the \bes{} population synthesis Galactic model that provides good agreement with stellar kinematics observed by HST towards the bulge. \mabuls{}-2 provides high-fidelity signal-to-noise limited maps of the microlensing optical depth, rate and average timescale towards a 400~deg$^2$ region of the Galactic bulge in several optical to near-infrared pass-bands. The maps take full account of the unresolved stellar background as well as limb-darkened source profiles. Comparing \mabuls{}-2 to the efficiency-corrected OGLE-IV 8,000 event sample shows a much improved agreement over the previous version of \mabuls{}, and succeeds in matching even small-scale structural features in the OGLE-IV event rate map. However, there remains evidence for a small under-prediction in the event rate per source and over-prediction in timescale. \mabuls{}-2 is available online (\url{www.mabuls.net}) to provide on-the-fly maps for user supplied cuts in survey magnitude, event timescale and relative proper motion.
\end{abstract}


%%%%%%%%%%%%%%%%%%%%%%%%%%%%%%%%%%%%%%%%%%%%%%%%%%

%%%%%%%%%%%%%%%%% BODY OF PAPER %%%%%%%%%%%%%%%%%%

\section{Introduction}

Gravitational microlensing is an important tool in exoplanet science, courtesy of its ability to detect exoplanets beyond the snowline \citep{Batista18}; an under-surveyed portion of the orbital parameter space that is important for testing planet formation theories \citep{Chunhua2013}.

Microlensing has also proven to be an important tool in the analysis of Galactic structure \citep{Moniez10} due to its ability to detect faint and low-mass objects such as M-dwarf stars and brown dwarfs that are typically undetectable beyond a few kiloparsecs. Accurate Galactic microlensing models also offer the ability to provide important prior constraints on the modelling of individual lens systems.

Theoretical models of microlensing optical depth $\tau$, Einstein crossing timescale $\displaystyle \left \langle t_{\rm E} \right \rangle$ or microlensing rate $\Gamma$ can be calibrated against efficiency-corrected observational data from ongoing large-scale microlensing surveys such as as MOA-2 \citep{Sumi2010}, OGLE-IV \citep{Udalski2015} and KMTNet \citep{Park2018}. The dependency of $\tau$ on the density distribution of the Galaxy makes this a good probe of Galactic structure, allowing us to compare empirical results such as from \cite{Mroz19} to simulation results. The microlensing rate $\Gamma$ probes not just the Galactic density distribution but also the lens and source kinematics and the lens mass function. It is therefore an important tool when informing future exoplanet microlensing surveys, such as the Nancy Grace Roman Space Telescope (hereafter NGRST, formerly WFIRST) \citep{Bennett2018} and the Vera Rubin Observatory (formerly LSST) \citep{Street18}, as regions of high $\Gamma$ will naturally yield more microlensing events and subsequently exoplanet detections. Finally, the Einstein radius crossing time $\displaystyle \left \langle t_{\rm E} \right \rangle$ is the characteristic timescale of a microlensing event and can be extracted directly from microlensing light curves. Measuring the timescale distribution from a statistically large sample of light curves can provide a good probe for estimating the mass distribution of lenses. Such data was used to calibrate a brown dwarf initial mass function (IMF) for microlensing simulations by \cite{Awiphan16} as well as providing tentative evidence for possible populations of free-floating planets (FFPs) \citep{Mroz17}.

Generating parameter maps from microlensing observations is difficult due to the requirement of a detection efficiency for events of a given timescale, which typically requires extensive Monte-Carlo simulations \citep{Sumi11}. Survey sky coverage, sampling rate and lifetime are also major limiting factors when generating empirical microlensing maps. In the work by \cite{Sumi13}, maps with resolutions of 1~degree were generated using around 470 events from the MOA-II survey. More recently, the OGLE-IV survey \citep{Mroz19} used around 8000 events to generate microlensing maps with a resolution of 10 arcminutes. Both analyses also considered two subsets of their microlensing data, namely red-clump giant (RCG) sources, which are bright and easily resolved, and difference image analysis (DIA) sources, which include fainter strongly blended source stars that may only be visible close to the magnification peak. Each of these subsets provide separate, but strongly correlated, measures of $\tau$.

The most detailed study to date comparing theoretical microlensing models and observational data was presented by \cite{Awiphan16} using the Manchester--\bes{} Microlensing Simulator (\mabuls{}\footnote{\url{http://www.mabuls.net/}}). \mabuls{} used the \bes{} Galactic populations synthesis model \citep{Robin2012B} to produce detailed microlensing maps for different photometric bands. All of the required ingredients for microlensing rate calculations, including lens mass, kinematic and density distributions, as well as source magnitude, kinematic and density distributions, are supplied by the \bes{} model, together with a fully calibrated 3D extinction model \citep{Marshall2006}. Comparison of \mabuls{} with MOA-II microlensing results reported evidence for a mass deficit in the Galactic bar as the model significantly under-predicted the observed optical depth. However \cite{Sumi2016} subsequently found a systematic problem with the way in which source stars were counted by MOA-2 in constructing their observed maps. Correcting for this, Penny \& Sumi found reasonable agreement between MOA-2 data and \mabuls{} predictions.

While our first generation \mabuls{} simulation provides a good level of agreement with MOA-2 data, the more recent OGLE-IV dataset provides a much greater challenge, involving a sample that is 17 times larger and allowing much higher fidelity observational maps. To provide a realistic comparison to such large datasets, we have developed a new generation of \mabuls{}. \mabuls{}-2 incorporates several important improvements, including calculations for fixed signal-to-noise ratio, accounting explicitly for unresolved stellar backgrounds of the event signal-to-noise, and a more detailed treatment of finite source size effects that are important for low mass FFPs. There is some independent evidence for a nearby FFP counterpart populations \citep{Liu13}, as well as individual candidates discovered through microlensing \citep{OGLE19}, so this will be a high priority area of study for future surveys such as NGRST. Whilst the previous version of \mabuls{} could calculate rates and optical depths for arbitrary magnitude and timescale cuts, \mabuls{}-2 additionally allows for arbitrary cuts in lens--source relative proper motion. This functionality makes it much more useful for providing model constraints on individual events where proper motion measurements are available.

The paper is structured as follows: Section~\ref{section:bgm} outlines the Galactic model used to simulate the microlensing sources and lenses used in this work, including the stellar kinematics and mass functions. Section~\ref{section:method} details the method behind generating the microlensing $\tau$, $\langle t_{\rm E} \rangle$ and $\Gamma$ maps, including the equations used for calculating these parameters from a discreet stellar population and lists the first order effects introduced to the simulation in this work, such as finite source effects and background light contributions. The results and discussion of the simulation are shown in section~\ref{section:results}, with microlensing parameter maps generated with different constraints, as well as a comparison with the OGLE-IV data compiled by \cite{Mroz19}. Conclusive remarks are given in section~\ref{section:conclusion}.

\section{The Besan\c{C}on Galactic Model}\label{section:bgm}

The approach of using Galactic population synthesis simulations to model microlensing was first demonstrated by \cite{kerins09} using the \bes{} Galactic Model (BGM) \citep{Robin03,Marshall2006,Robin2012B}. This method has also been used to make detailed predictions for exoplanet microlensing yields for forthcoming space missions such as Euclid \citep{penny13} and NGRST \citep{Penny19} using a more recent version of the BGM (variant BGM1106). Recently, a population synthesis based microlensing model has been developed to study microlensing due to black holes \citep{popsycle}.

\cite{Penny19} noted that the BGM1106 model does not provide a close match to the HST stellar kinematics study of \cite{Clarkson08}. \cite{Awiphan16} presented a more recent version of the model (variant BGM1307) and compared it to a large ensemble of 470 events from the MOA-2 survey \citep{Sumi13}. \cite{Awiphan16} concluded that there was a reasonable level of agreement once allowance was made for brown dwarfs that are missing from the BGM. However, the model appeared to significantly under-predict the observed rate from the Galactic bulge. \cite{Sumi2016} subsequently showed that at least part of the cause of this discrepancy was due to under-counting of source stars within the MOA-2 analysis, and correcting for this resulted in much better agreement with the model. After investigation of a number of different BGM variants we have decided to use BGM1307 as the basis for \mabuls{}-2.

The BGM separates the Galaxy into four components; the thin disk, thick disk, bar and halo, each with their own stellar initial mass functions (IMFs), star formation rates (SFRs), kinematics and ages. Interstellar extinction is included for $UBVRIJHK$ photometric bands using a 3D extinction model from \cite{Marshall2006}. The solar position relative to the Galactic centre and plane in the BGM is taken to be $R_0 = 8$~kpc and $z_0 = 15$~pc, respectively.

The following description of the Galactic components quotes lower limits on their respective stellar mass functions, but an upper limit is not included. While interesting lens candidates such as intermediate mass black holes and other compact objects are represented in the high mass regime, the BGM does not include them. Due to the steepness of the mass functions at high mass, these objects also do not contribute significantly to the overall microlensing parameters outlined in section \ref{section:parameters}.

\subsection{The Thin Disk}

The thin disk component is sub-divided into seven age groups, each with their own luminosity and effective temperature distributions. The mass density is modelled by the difference of two exponentials in cylindrical polar coordinates \textit{(r, z)}; one representing the body of the disk with scale length $R_{d} = 2170$~pc and the other representing a central hole with scale length $R_{h} = 1330$~pc \citep{Robin2012A}. It is given by
\begin{equation}
\rho(r, z) = \rho_{0}\Bigg\{exp\Bigg(-\sqrt{\frac{1}{4} + \frac{a^2}{R_{d}^2}}\Bigg) - exp\Bigg(-\sqrt{\frac{1}{4} + \frac{a^2}{R_{h}^2}}\Bigg) \Bigg\},
\end{equation}

where $\rho_{0}$ is the component normalisation density, $a^2 = r^2+\big(\nicefrac{\displaystyle z}{\displaystyle \epsilon}\big)^2$ and $\epsilon$ the axis ratio of the ellipsoid. The SFR is taken to be constant across the whole disk. The IMF is given by a broken power law, which is the same across all age groups

\begin{equation}
\zeta(M) \propto \begin{cases}
               M^{-1.6},\hspace{8.5mm} 0.079M_\odot\ \leq M < 1M_\odot\\
               M^{-3},  \hspace{10mm} M \geq 1M_\odot\\
            \end{cases}.    
\end{equation}
Each age group of the thin disk has its own velocity dispersion, to mimic the secular evolution, and follows the relation from \cite{Gomez97}; these average to $\mathbf{\sigma_{UVW}} = (30,20,13)$~\kms, with a maximum of $\mathbf{\sigma_{UVW}} = (43,28,18)$~km~$s^{-1}$ for the oldest component. The rotational velocity of the local standard of rest (LSR) is  240~\kms, with the rotation curve taken from \cite{Caldwell81}. The LSR is defined as the mean velocity of material in the solar neighbourhood, centred on the Sun within 100~pc. The peculiar motion of the sun relative to the LSR is assumed to be $\mathbf{v_\odot}=(11,12,7)$~\kms, following \cite{Schonrich10}.

\subsection{The Thick Disk}

The thick disk is an older and less dense component of the model. It is most prevalent at latitudes beyond the scope of the thin disk at $|b| > 9^{\circ}$ and has a modified exponential density law with scale height $h_{z} = 533.4$~pc, scale length $h_{R} = 2355.4$~pc and break distance $\xi = 658$~pc,
\begin{equation}
    \rho(r, z) = \rho_{0}exp\Bigg(\frac{r_{\odot}-r}{h_{R}}\Bigg)\begin{cases}
               1 - \Bigg(\frac{z^2}{\xi(2h_{Z}+\xi)}\Bigg),\hspace{12.3mm} z \leq \xi\\
               \frac{2h_{z}}{2h_{z}+\xi}exp\Bigg(\frac{\xi-|z-z_{\odot}|}{h_{z}}\Bigg),  \hspace{5mm} z > \xi\\
            \end{cases},
\end{equation}
where $\rho_{0}$ is the local stellar density with $\big(r_{\odot},z_{\odot}\big)$ representing the coordinates of the sun. The star formation history is modelled by a single burst 10 Gyr ago \citep{Robin14}. We use the \cite{Bergbusch92} luminosity function with an isochrone of 10 Gyr, a mean metallicity of -0.5 dex and alpha enhanced to simulate this population. The IMF of the thick disk is given by a single power law of the form
\begin{equation}
    \zeta(M) \propto M^{-1.5}, \hspace{10mm} M > 0.154M_{\odot}.
\end{equation}
The velocity dispersion of the thick disk is taken to be $\mathbf{\sigma_{UVW}} = (67,51,42)$~\kms with a rotational velocity of 176~\kms.

\subsection{The Bar}\label{section:bar}

The bar population dominates Galactic densities at low latitudes, $|b| < 5^{\circ}$. The bar is represented by a Padova isochrone of 8 Gyr, with a solar metallicity. It is modelled as a triaxial ellipsoid with scale lengths $x_{0} = 1.46$~kpc, $y_{0} = 0.49$~kpc, $z_{0} = 0.3$~kpc, with the major axis (\textit{x}) offset from the sun-Galactic centre axis by $12.89^{\circ}$ \citep{Robin2012A}. Its pitch and roll is set to be zero. Its density distribution is given in galactocentric cartesian coordinates as
\begin{equation}
    \rho(x,y,z) = \rho_{0}sech^2\big(-R_{s}(x,y,z)\big),
\end{equation}
\begin{equation}
    R_{s}(x,y,z)^{C_{\parallel}} = \Bigg(\Bigg[\frac{x}{x_{0}}\Bigg]^{C_{\perp}} +\hspace{2mm} \Bigg[\frac{y}{y_{0}}\Bigg]^{C_{\perp}} \Bigg)^\frac{C_{\parallel}}{C_{\perp}} + \Bigg(\frac{z}{z_{0}}\Bigg)^{C_{\parallel}}.
\end{equation}
The bar density function is then multiplied by a gaussian cutoff function of width 0.5~kpc in the xy plane to confine the radius $R_{xy} = \sqrt{x^2 + y^2}$ within the cutoff radius $R_C =$ 3.43~kpc,
\begin{equation}
    f_c(R_{xy}) = \begin{cases} 1, \hspace{30.5mm} R_{xy} \leq R_C \\
    exp\Bigg(-\Big(\frac{R_{xy} - R_C}{0.5}\Big)^2\Bigg), \hspace{5.0mm} R_{xy} > R_C
    \end{cases}
\end{equation}
The parameters $C_{\parallel}$ and $C_{\perp}$ control the 'disky' or 'boxy' shape of the ellipsoid; for this version of the BGM, a boxy bar shape was used, with $C_{\parallel} = 0.5$ and $C_{\perp} = 3.007$. The bar IMF was modelled with a broken power law,
\begin{equation}
\zeta(M) \propto \begin{cases}
               M^{-1.5},\hspace{11.5mm} 0.15M_\odot\ \leq M < 0.7M_\odot\\
               M^{-2.35},\hspace{10mm} M \geq 0.7M_\odot\\
            \end{cases}.       
\end{equation}
The bar has the largest velocity dispersion with $\mathbf{\sigma_{UVW}} = (150,115,100)$~\kms. Its kinematics have been modelled by an N-body simulation, outlined in \cite{Gardner2014}.

\subsection{The Halo}

The halo component mainly contributes at high latitudes and for magnitudes fainter than $m\sim18$. Stars in the halo are generated with an age of 14 billion years and naturally have a low metallicity, with [Fe/H] = -1.78 \citep{Robin14}. The density profile of the halo is given by a power law,
\begin{equation}
    \rho(r,z) =   \rho_{0}\Bigg(r^2+\bigg(\frac{\displaystyle z}{\displaystyle \epsilon}\bigg)^2\Bigg)^{-1.695},
\end{equation}
where $\epsilon = 0.768$ is the axis ratio of the ellipsoid. The halo IMF follows a single power law,
\begin{equation}
    \zeta(M) \propto M^{-1.5}, \hspace{10mm} M > 0.085M_{\odot}.
\end{equation}
The halo has a velocity dispersion of $\mathbf{\sigma_{UVW}} = (131,106,85)$~\kms.

\begin{table}
\caption{Summary of the dispersion velocities for each component in BGM1307, using the galactic (UVW) coordinate system. The numbers listed for the thin disk are averaged over each age group sub-division.}
\label{table:table0}
\centering
\begin{tabular}{ C{1.7cm} C{1.65cm} C{1.65cm} C{1.65cm} } 
    \hline\hline
    \textbf{Component} & $\sigma_U$~\kms & $\sigma_V$~\kms & $\sigma_W$~\kms \\ 
    \hline
    Thin Disk & 30 & 20 & 13 \\ 
    Thick Disk & 67 & 51 & 42 \\ 
    Bar & 150 & 115 & 100 \\ 
    Halo & 131 & 106 & 85 \\ 
    \hline
 \hline
\end{tabular}
\end{table}

\subsection{Comparison with empirical data}

\begin{figure*}
    \centering
    \includegraphics[width=12cm]{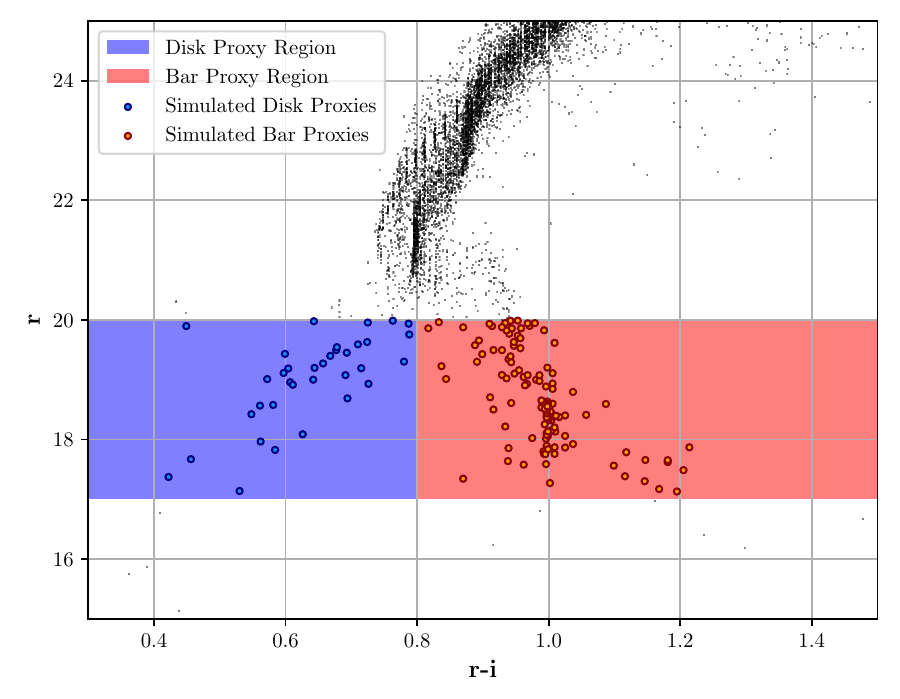}
    \caption{The colour magnitude diagram of the simulated source stars from BGM1307. The selection of the sloan $r$ and $i$ colours were chosen to represent the Hubble F814W and F606W filters. The stellar distribution bifurcates at around $r=20$ into the two separate blue disk and red bar proxy populations, with the partition between them placed at $r-i=0.8$. The disk and bar selection regions are highlighted as the blue and red rectangles, while the proxy stars themselves are shown as outlined blue or orange circles, respectively. The dark dots above the proxy regions are sources which do not act as proxy stars, but were still generated by the BGM synthesis code.}
    \label{figure:proxy_select}
\end{figure*}

\begin{figure*}
    \centering
    \includegraphics[width=15cm]{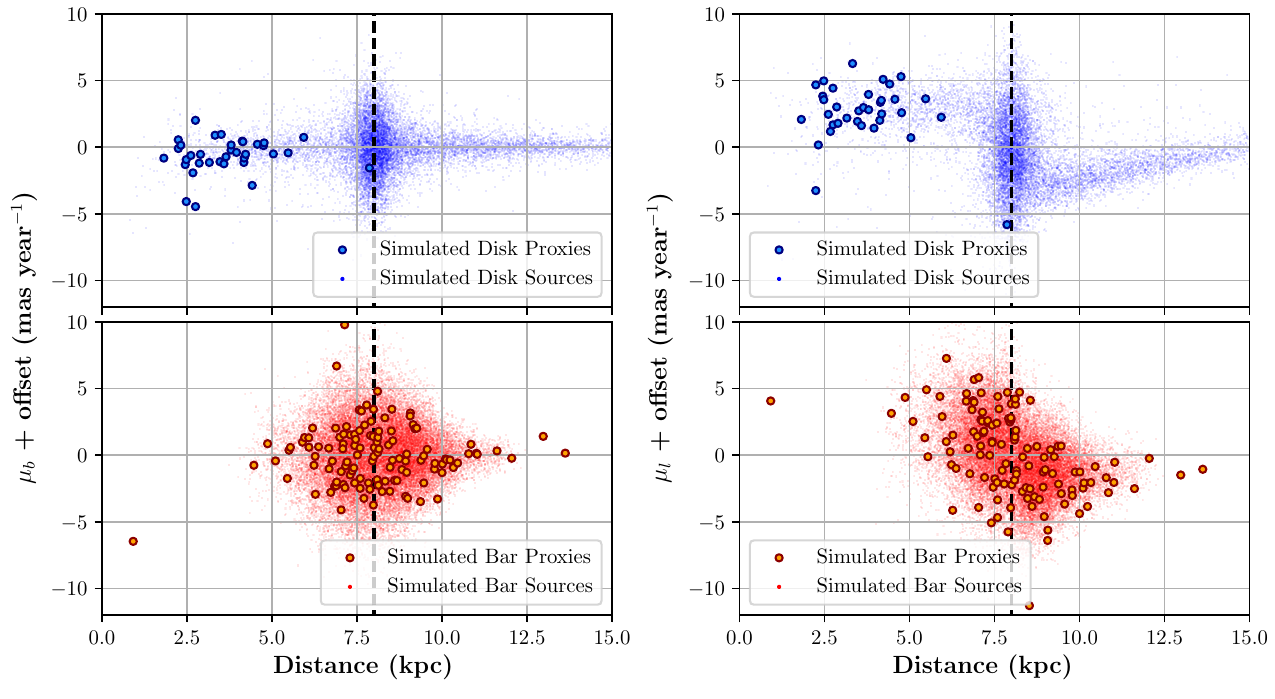}
    \caption{Shown are the proper motion vs. source distance distributions for the synthetic disk and bar populations. Proxy stars are shown as solid, outlined circles, corresponding to the proxy stars in figure \ref{figure:proxy_select}. The proper motion axes are indicated with an offset; this is because the HST data was calculated relative to the bar proxy population's proper motion, hence centering the bar population on $(\mu_{\rm l},\mu_{\rm b})=(0,0)$~\msyr. On the top left is the distribution of proper motion in Galactic latitude, with the distance to the Galactic centre indicated with the black dashed line at 8~kpc. The small blue dots are source stars which were simulated as part of the Galactic disk (either thin or thick disk), but were not part of the proxy population. The bottom left shows the same, but for the bar proxies and sources. The right figures show the same as the left figures, but for the distribution of proper motion in Galactic longitude.}
    \label{figure:star_dist}
\end{figure*}

\begin{figure*}
    \centering
    \includegraphics[width=15cm]{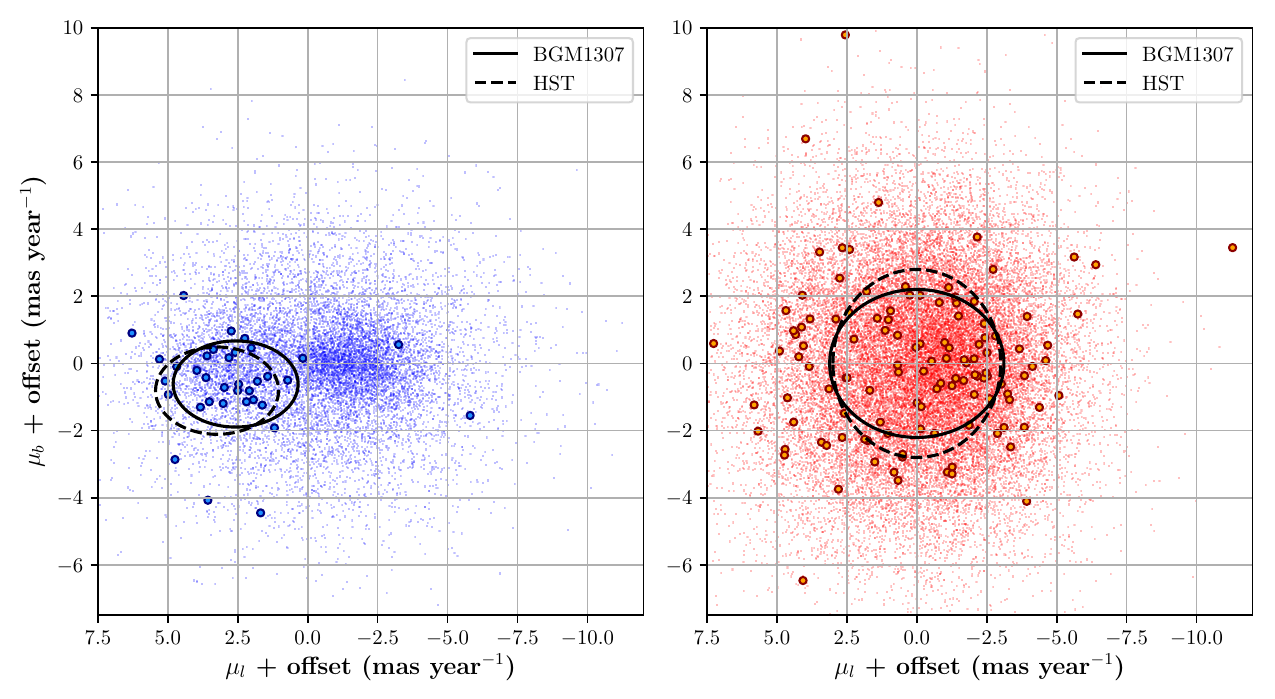}
    \caption{Shown are the proper motion distributions in both Galactic latitude and longitude for the disk proxies (left) and bar proxies (right). The proxy and source markers follow the same convention as figure \ref{figure:star_dist}. The $1\sigma$ contours for the simulated proxy populations are shown as solid black ellipses. The equivalent contours for the HST proxy stars are shown as dashed ellipses. Due to the offset mentioned in figure \ref{figure:star_dist}, both bar contours are centred on $(\mu_{\rm l},\mu_{\rm b})=(0,0)$~\msyr.The BGM bar contour has a width of $(\sigma_l,\sigma_b) = (3.12,2.2)$~\msyr, while the HST contour has a width of $(\sigma_l,\sigma_b) = (3.0,2.8)$~\msyr. The BGM disk contour is centred on $(\mu_l,\mu_b) = (2.58,-0.61)$~\msyr~ and width $(\sigma_l,\sigma_b) = (2.23,1.28)$~\msyr. The HST disk contour is centred on $(\mu_l,\mu_b) = (3.24,-0.81)$~\msyr~ and width $(\sigma_l,\sigma_b) = (2.2,1.3)$~\msyr.}
    \label{figure:proper_motion}
\end{figure*}

Figures~\ref{figure:proxy_select}, \ref{figure:star_dist} and \ref{figure:proper_motion} follow the kinematics comparison of BGM1106 by \cite{Penny19}, this time applied to BGM1307. The model predictions are compared to stellar kinematics derived from HST data presented by \cite{Clarkson08}. This is achieved by generating and selecting proxy stars from BGM1307 which closely match the selection criteria of the HST disk and bar populations. The alignment of the $1\sigma$ contours shows very good agreement between BGM1307 predicted and HST observed kinematics and is a clear improvement over BGM1106. The kinematics for BGM1307 are summarised in table \ref{table:table0}. Comparing the BGM to Gaia data in \citet{Arenou2017} shows an over-prediction by the model of the star counts, attributed partly to the incompleteness of the Gaia catalogues for the faintest stars, but also showing an over-prediction for the brightest stars in the bulge region around $|b| < 2^\circ$, attributed to inaccuracies in the extinction model. For comparison with the OGLE-IV survey, this is less of an issue, as the survey has little data in the $|b| < 1^\circ$ region. Outside of the bulge, between $2^\circ < |b| < 10^\circ$, there is good agreement between the Galactic model and the Gaia data.

\subsection{Low mass stars and brown dwarfs}

\begin{figure}
\includegraphics[width=8cm]{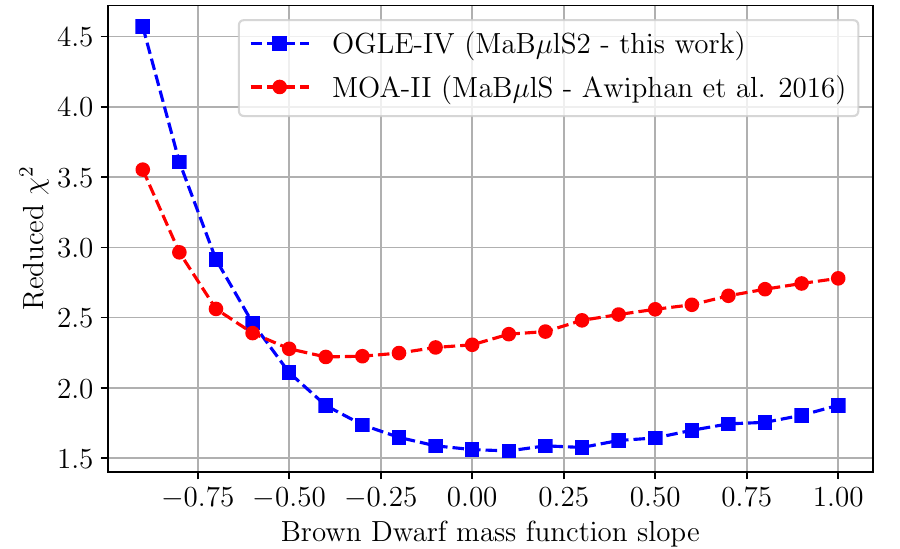}
\caption{The $\chi^2$ minimisation of the brown dwarf mass function slope. The optimal value of 0.1 is notably different from the previous version, optimised with MOA II data by \citet{Awiphan16}, which yielded an optimal slope of -0.4.}
\label{figure:figureBD}
\end{figure}

Low mass M~dwarfs are not included in some components of the BGM, as evidenced by some of the IMF mass limits discussed above. Brown dwarfs are also not included in the model. To compensate for this, additional red dwarfs were added, down to the hydrogen burning limit for the thick disk, bar and halo, by continuing their respective IMFs. Their kinematics and distances were inherited at random from synthetic stars from catalogues of any of the 4 magnitude ranges for a particular line of sight. Due to their low luminosities, they were assigned an apparent magnitude of 99, excluding them as potential sources. Brown dwarfs and high mass FFPs were also added with an apparent magnitude of 99 and a mass range between the hydrogen burning limit and $\sim M_{\jupiter}$  using an IMF
\begin{equation}
    \zeta(M) \propto M^{0.1}, \hspace{10mm} 0.001M_{\odot} < M < 0.079M_{\odot}.
\end{equation}
This IMF slope was acquired by performing a $\chi^2$ minimisation over a range of slope values in the interval $[-0.9,1.0]$ in 0.1 increments, as shown in figure \ref{figure:figureBD}. The minimisation used 258 evenly spaced points in the OGLE timescale map by \cite{Mroz19}. A modified version of the \mabuls{} algorithm, outlined in section \ref{section:ogledata}, was used to ensure event selection criteria were as faithful as possible to the OGLE events. Although this work is comparing the \mabuls{}-2 prediction of $\langle t_{\rm E} \rangle$ to the data after fitting a brown dwarf mass slope to minimize the $\chi^2$ between the two, it is worthy of note that the optical depth is not significantly affected by the addition of brown dwarfs and still matches the data quite well (section \ref{section:ogledata}). The previous version of \mabuls{} also attempted this minimization on MOA II timescales, but does not achieve as successful a result as the new simulation, as shown in figure \ref{figure:figureBD}. As the OGLE data is being used to extract a brown dwarf mass function and hence the shortest event timescales, we are only testing the longer timescale regime.

\begin{table}
\caption{The range of apparent magnitude in each magnitude interval for simulated stars is shown. Low mass objects from section 2.5 only appear in interval 4 as they are assigned an apparent magnitude of 99. Microlensing sources are only considered from intervals 1 $\rightarrow$ 3.}
\label{table:table1}
\centering
\begin{tabular}{ C{3cm} C{3cm} } 
    \hline\hline
    \textbf{Interval No.} & \textbf{Magnitude Range} \\ 
    \hline
    1 & $0 \leq K < 15$ \\ 
    2 & $15 \leq K < 20$ \\ 
    3 & $20 \leq K < 24$ \\ 
    4 & $24 \leq K < 99$ \\ 
    \hline
 \hline
\end{tabular}
\end{table}

\subsection{Synthetic catalogue sizes}

The BGM allows the user to specify the Galactic coordinates of the simulation line of sight, as well as the solid angle in deg$^2$ over which to simulate stars. The simulation volume is thus a square based pyramid, with an opening angle equivalent to the square root of the solid angle. The size of the solid angle is used to cap the number of stars to fit computational requirements. However, we must also ensure that we sample rarer bright stars sufficiently well. To this end, for each line of sight, four catalogues were generated, each covering a fixed range of apparent magnitudes shown in Table \ref{table:table1}. The magnitude ranges were chosen to sample different portions of the mass functions; high mass, low occurrence stars have much higher weights as source stars, while low mass high occurrence stars dominate the lensing contribution. The solid angles of each magnitude interval for each line of sight were calibrated to produce $\sim$ 10,000 stars per catalogue resulting in 40,000 stars in total, per line of sight. This total of 40,000 stars was chosen to control the computation time of the simulation code; allowing for constant solid angles per magnitude range results in a drastic and unnecessary increase in computation time along dense lines of sight, namely the Galactic bar, for diminishing returns in error reduction. Our microlensing calculations must therefore be appropriately re-scaled to account for these solid angle choices.

\section{Simulation method}\label{section:method}

\subsection{Microlensing parameters}\label{section:parameters}

The simplest gravitational microlensing scenario is the point-source point-lens (PSPL or Paczynski) model \citep{Paczynski86}. In this case, both the lens and source are considered single objects with an infinitesimal angular size; this is accurate to describe most microlensing events observed to date. The magnification due to the lens as a function of time $A(t)$ at a specific normalised angular impact parameter $u(t)$, is given by
\begin{equation}
    \label{equation:equation11}
    A(t) = \frac{u(t)^2 + 2}{u(t)\sqrt{u(t)^2 + 4}},
\end{equation}
\begin{equation}
    u(t)^2 = u_{0}^2 + \Bigg(\frac{t-t_{0}}{t_{\rm E}}\Bigg)^2,
\end{equation}
where $t_{0}$ is the time of maximum magnification and $u_{0}$ is the minimum impact parameter. The impact parameter is normalised to the angular Einstein radius,
\begin{equation}
    \theta_{\rm E}=\sqrt{\frac{4GM}{c^2}\frac{D_{\rm s}-D_{\rm l}}{D_{\rm s}D_{\rm l}}},\hspace{10mm}D_{\rm s}>D_{\rm l},
\end{equation}
for a lens mass $M$ at a distance $D_{\rm l}$ and a source at a distance $D_{\rm s}$. The microlensing optical depth $\tau$ is
\begin{equation}
    \tau = \frac{4\pi G}{D_{\rm s}c^2}\int_{0}^{D_{\rm s}}{\rho(D_{\rm l})D_{\rm l}(D_{\rm s}-D_{\rm l})dD_{\rm l}},
\end{equation}
for a continuous lens mass-density distribution $\rho(D_{\rm l})$. For a discrete catalogue of $N_{s}$ source stars and $N_{l}$ lens stars we can instead use \citep{kerins09,Awiphan16}:
\begin{equation}
    \label{equation:equation15}
    \tau = \pi\frac{\sum_{s}^{N_{s}}\sum_{l}^{N_{l};D_{\rm s}>D_{\rm l}}u_{\max}^2\theta_{\rm E}^2\frac{1}{\Omega_s\Omega_l}}{\sum_{s}^{N_{s}}\langle w^2 \rangle \frac{1}{\Omega_s}}.
\end{equation}
The variable $u_{\max} = u(A_{\min})$ represents the largest impact parameter that permits a magnification $A \geq A_{\min}$. It is common to adopt an absolute threshold maximum impact parameter $u_{\rm t}$, where often $u_{\rm t} = 1$ is adopted. We can incorporate this into the definition of $u_{\max}$ by defining $u_{\max} = min[u(A_{\min}),u_{\rm t}]$. The variable $\langle w^2 \rangle$ is the 2$^{nd}$ moment of the source weight $w$, which counts the effective number of sources; in the case of a source resolved by the telescope at baseline, this will be equal to unity, however for DIA sources which are only visible during magnification, $w$ can drop below unity, resulting in a smaller contribution to the total optical depth. In general, the p$^{th}$ moment of $w$ is given by
\begin{equation}
    \langle w^p \rangle = \frac{\sum_{l}^{N_{l}}w^p\mu_{\rm rel}D_{\rm l}^2\theta_{\rm E}\frac{1}{\Omega_l}}{\sum_{l}^{N_{l}}\mu_{\rm rel}D_{\rm l}^2\theta_{\rm E}\frac{1}{\Omega_l}},
\end{equation}
where $\mu_{\rm rel}$ is the lens-source relative proper motion. One way to consider equation \ref{equation:equation15} is as a summation over the 'sensitivity regions' in the sky formed by circles of radius $u_{\max}\theta_{\rm E}$, averaged over all possible sources, as a ratio to the total survey solid angle. The rate-weighted average Einstein radius crossing time $\langle t_{\rm E} \rangle$ for a line of sight is given by
\begin{equation}
\label{equation:equation17}
    \langle t_{\rm E} \rangle = \frac{\sum_{s}^{N_{s}}\sum_{l}^{N_{l};D_{\rm s}>D_{\rm l}}wD_{\rm l}^2\theta_{\rm E}^2\frac{1}{\Omega_s\Omega_l}}{\sum_{s}^{N_{s}}\sum_{l}^{N_{l};D_{\rm s}>D_{\rm l}}w\mu_{\rm rel}D_{\rm l}^2\theta_{\rm E}\frac{1}{\Omega_s\Omega_l}}.
\end{equation}
Similarly, the rate-weighted average relative proper motion $\langle \mu_{\rm rel} \rangle$ for a line of sight is given by
\begin{equation}
    \langle \mu_{\rm rel} \rangle = \frac{\sum_{s}^{N_{s}}\sum_{l}^{N_{l};D_{\rm s}>D_{\rm l}}w\mu_{\rm rel}^2D_{\rm l}^2\theta_{\rm E}\frac{1}{\Omega_s\Omega_l}}{\sum_{s}^{N_{s}}\sum_{l}^{N_{l};D_{\rm s}>D_{\rm l}}w\mu_{\rm rel}D_{\rm l}^2\theta_{\rm E}\frac{1}{\Omega_s\Omega_l}}.
\end{equation}
Finally, the microlensing rate per source star is simply given by
\begin{equation}
    \Gamma = \frac{2}{\pi}\frac{\tau}{\langle u_{\rm t}t_{\rm E} \rangle}.
\end{equation}

\begin{figure}
\includegraphics[width=8cm]{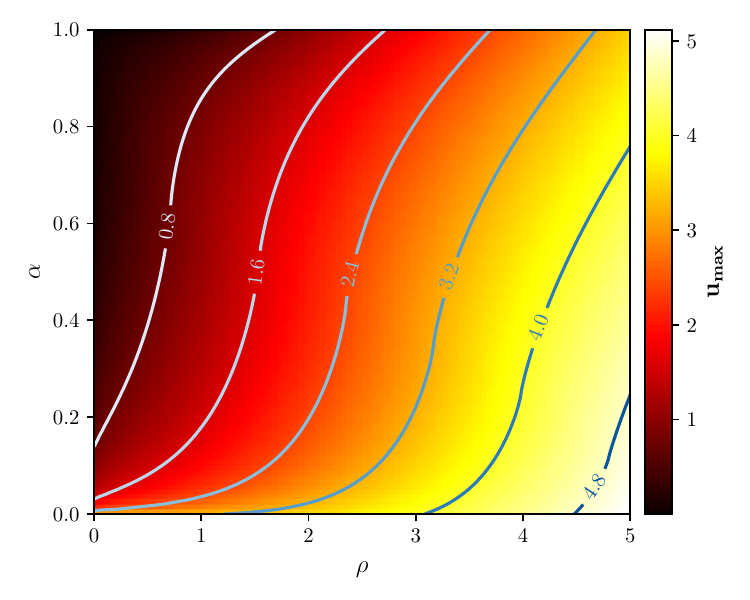}
\caption{The $u_{\max}$ distribution as a function of normalised source radius $\rho$ and magnification threshold proxy $\alpha$ is shown, with $u_{\max}$ contours. The top of the distribution ($\alpha = 1$) represents the cutoff magnification from equation \ref{equation:equation20} which has been re-scaled using equation \ref{equation:equation21} to more efficiently use the space available. The logarithmic scaling also stretches/compacts the distribution}
\label{figure:figure1}
\end{figure}

Note, that while we are drawing both lenses and sources from the same catalogues, we are demanding that any lens objects must satisfy $D_s > D_l$ for the event to occur. This is handled by the code during the calculation of all microlensing parameters by skipping over lens candidates in the catalogue which do not satisfy this condition while calculating the contribution of a particular source star. Similarly, not all stars in the catalogue are valid source stars due to their apparent magnitude and are thus skipped by the code during calculation.

\subsection{Finite source effects}

In cases where the angular size of the source star is comparable to its impact parameter with the lens, finite source effects may become evident in the light curve, as different parts of the source are magnified by appreciably different amounts. This typically results in a flattening of the light curve peak, as the singularity in equation \ref{equation:equation11} is avoided as $u \rightarrow 0$. This has an important consequence for the optical depth, as $u_{\max} = u(A_{\min})$ can no longer be inverted from equation \ref{equation:equation11}, but requires a numerical solution. The treatment by \cite{Lee09} was used to evaluate $u_{\max}$ as a function of both $A_{\min}$ and the angular source radius normalised to the Einstein radius, $\rho$. A lookup table was generated with a resolution of $100 \times 100$ $u_{\max}$ evaluations over the range $0.01 \leq \rho \leq 5$ and $ A_{\rm base} \leq A_{\min} \leq 200$, with $A_{\rm base} = 1.01695$ (the PSPL magnification for $u=3$); bilinear interpolation was then used to extract a continuous value for use in the simulation.

For a uniform disk, there is a maximum possible magnification $A_{\rm cut}$ which a microlensing event can reach for a particular $\rho$ given by
\begin{equation}
\label{equation:equation20}
    A_{\rm cut} = A(\rho,u=0) = \sqrt{1 + \frac{4}{\rho^2}}.
\end{equation}
 Consequently, simply using $A_{\min}$ for the vertical axis of the $u_{\max}$ distribution results in most of the grid being zero due to finite source effects preventing magnifications greater than $A_{\rm cut}$ from being achieved. To account for this, a magnification threshold proxy parameter $\alpha$ was used, where
\begin{equation}
\label{equation:equation21}
    \alpha = \frac{\ln(A_{\min} - A_{\rm base} + 1)}{\ln(A_{\rm cut} - A_{\rm base} + 1)},
\end{equation}
which scales the top of the distribution to $A_{\rm cut}$ and also more evenly distributes the information, resulting in a more slowly varying derivative, as evident from figure \ref{figure:figure1}.
\newline\indent As $\rho \rightarrow 1$, stellar limb darkening becomes a significant factor. To deal with this, a linear limb darkening coefficient was introduced to the numerical calculation, which varies as a function of observation wavelength and effective temperature. $UBVRIJHK$ linear LDCs from \cite{Claret11} were used to construct curves in each wavelength band (excluding $L$ band, which reused $K$-band data) as a function of effective temperature, which were then interpolated in real time during the simulation for each source star. To account for this in the $u_{\max}$ grid, a simple scaling of $\alpha$ and $\rho$ were performed before the bilinear interpolation to approximate the effects of limb darkening. This was necessary to reduce computation time, as no analytical version of equation \ref{equation:equation20} exists with limb-darkening considerations.

\subsection{Background light contributions}\label{section:bg_light}

The calculation of $A_{\min}$ is derived from the event selection criterion that at peak magnification, a signal-to-noise ratio $S/N \geq 50$ must be achieved. In the previous version of \mabuls{}, calculations were based only on a source magnitude threshold cut, not a survey $S/N$ cut. For our survey $S/N$ cut we assume three component contributions: photons from the source $N_{\rm src} = t_{\rm exp}A_{\min}10^{-0.4(m_{\rm s}-m_{\rm zp})}$, a uniform sky background $N_{\rm sky} = t_{\rm exp}\Omega_{\rm psf}10^{-0.4(\mu_{\rm sky}-m_{\rm zp})}$ and the light contribution from all other stars under the source's PSF, $N_{\rm BG} = t_{\rm exp}\Omega_{\rm psf}\sum_{i}10^{-0.4(m_{i}-m_{\rm zp})}\frac{1}{\Omega_{i}}$. The combination of the zero-point magnitude $m_{\rm zp}$ and exposure time $t_{\rm exp}$ is fixed to provide 4\% photometric precision for an assumed seeing of $\theta_{\rm FWHM} = 1$~arcsec, (giving $\Omega_{\rm psf} = 0.785$~arcsec$^2$) at a telescope limiting magnitude $m_{\rm lim}$ \citep{Ban16}. $4\%$ photometric precision is therefore achieved for $m_{\rm lim} = m_{\rm zp}$ if $t_{\rm exp} = 625$~s. More generally, the $S/N$ at peak magnification is
\begin{equation}
    \label{equation:equation22}
    S/N = \frac{t_{\rm exp}10^{-0.4(m_{\rm s}-m_{\rm zp})}A_{\min}}{\sqrt{N_{\rm BG} + N_{\rm sky} + t_{\rm exp}10^{-0.4(m_{\rm s}-m_{\rm zp})}A_{\min}}},
\end{equation}
with the corresponding value of $A_{\min}$ given by
\begin{equation}
    \label{equation:equation23}
    A_{\min} = \frac{(S/N)^210^{0.4m_{\rm s}}}{2 t_{\rm exp} 10^{0.4m_{\rm zp}}}\Bigg(1 + \sqrt{1 + \frac{4(N_{\rm BG} + N_{\rm sky})}{(S/N)^2}}\Bigg).
\end{equation}

The particular value of $\mu_{\rm sky}$ is band dependent. Ground-based values for Paranal Observatory from \citet{Patat2003} are adopted. These values are shown in table \ref{table:table_musky}. We take $t_{\rm exp} = 625$~s and use the survey limiting magnitude for $4\%$ photometric precision, $m_{\rm lim}$, for $m_{\rm zp}$. While a seeing of $\theta_{\rm FWHM} = 1$~arcsec is reasonable for a microlensing survey such as OGLE-IV, it is less representative of telescopes such as MOA's Mt. John Observatory, where due to atmospheric conditions, seeing is typically larger. To simulate this, the code was run to sample the optical depth and timescale along eight lines of sight near Baade's Window. We find that the optical depth under MOA-II conditions was found to be a factor 0.93 that for OGLE-IV, therefore broadly similar. We find no statistically significant difference between our predictions for the mean timescale of MOA-II and OGLE-IV.

\begin{table}
\caption{The sky brightness as a function of Johnson-Cousins filter.}
\label{table:table_musky}
\centering
\begin{tabular}{ C{2cm} C{4cm} } 
    \hline\hline
    \textbf{Filter} & \textbf{Sky Brightness / mag arcsec$^{-2}$} \\ 
    \hline
    U & 22.28 \\ 
    B & 22.64 \\ 
    V & 21.61 \\ 
    R & 20.87 \\ 
    I & 19.71 \\ 
    J & 16.50 \\ 
    H & 14.40 \\ 
    K & 13.00 \\ 
    \hline
 \hline
\end{tabular}
\end{table}

\subsection{Error calculation}

The treatment of the parameter errors has been made more rigorous; previously, the error was estimated by distributing all sources into two bins and taking the difference of the two results as an approximation for the error. This ignored the variance due to the lenses, which were kept constant across both bins, as well as the inaccuracies of estimating the standard deviation with only two source bins. To account for both shortcomings, ten bins were used for both sources and lenses. The error for a particular source, $\sigma_{s}$ was calculated by distributing the lenses randomly into the ten lens bins and calculating the standard deviation. At this stage in the error propagation, only the numerator terms of equations \ref{equation:equation15} and \ref{equation:equation17} were calculated, as the numerators and normalisation terms must be summed up separately during map generation. After obtaining the parameter values $x_{s}$, error estimates $\sigma_{s}$ and normalisation $\langle w_{s} \rangle$ for a particular source, the ten source bins were then populated by looping through all sources. The sums over $x_{s}$, $\sigma_{s}$ and $\langle w_{s} \rangle$ in source bin \textit{i} are $x_{i}$, $\sigma_{i}$ and $\langle w_{i} \rangle$ respectively. The final error value on a parameter \textit{x} is
\begin{equation}
    \epsilon = \sqrt{\frac{\sum_{i}(x_{i}^2+\sigma_{i}^2)\langle w_{i} \rangle}{\sum_{i}\langle w_{i} \rangle} + 
    \Bigg(\frac{\sum_{i}x_{i}\langle w_{i} \rangle}{\sum_{i}\langle w_{i} \rangle}\Bigg)^2}.
\end{equation}

\subsection{Map generation method}

The simulation distributes the resulting parameters, normalisations and errors into 3-dimensional bins of source star magnitude $m_{\rm s}$ (10 bins), average timescale $\langle t_{\rm E} \rangle$ (10 bins) and average relative proper motion $\langle \mu_{\rm rel} \rangle$ (5 bins). The bin edges for apparent magnitude were uniformly sampled between 12 and 23. For mean timescale, the 10$^{th}$ percentiles of the $I$-band timescale distribution multiplied by the timescale were found; the multiplication by timescale was implemented to account for the fact that event weights for optical depth are linearly proportional to their corresponding timescale. The bin edges for $\mu_{\rm rel}$ were found by calculating the 20$^{th}$ percentiles of the $I$-band distribution. 
\newline\indent The output files contain the integrals of optical depth and timescale up to their corresponding upper bin edges; this was to eliminate the necessity of looping over all bins interior to a user-specified cut in each of the three dimensions, allowing the map generator to simply add and subtract the integral limits. To account for a user specified cut in any direction which does not lie on bin edges (as would be the case most of the time), 6D interpolation is performed using the up-to 64 integrals bounding the user's selected parameter ranges; this is the product of 3D interpolation in the bin containing the upper bound and the bin containing the lower bound. The dimensionality of the interpolation method is automatically reduced if the user picks bounds on a bin edge.
\newline\indent This method of interpolation is more accurate than the previous version of \mabuls{}, which used 2-dimensional interpolation, using $m_{\rm s}$ and $\langle t_{\rm E} \rangle$ cuts, over the parameter values, errors and normalisations, before integrating over these interpolants. Switching the order of operations to interpolating over the integrals prevents inaccuracies accumulated by the interpolation process, as integrals confined to bin edges are exact; as such, interpolating over them ensures that the result will vary smoothly between the various bounds and is also more reliable at preserving timescale ranges to be within the range specified.

\section{Results}\label{section:results}

Examples of microlensing maps are shown below. The parameters $\tau$, $\langle t_{\rm E} \rangle$ and $\Gamma$ are simulated over the region $|b| \leq 10^{\circ} \cup |l| \leq 10^{\circ}$, centred on the Galactic coordinates origin, $l,b = 0^{\circ}$. The microlensing event rate $\Gamma$ is shown as either the event rate per star or the event rate per square degree. In figures \ref{figure:figure3}, \ref{figure:figure4} and \ref{figure:figure5}, these maps are shown in $V$, $I$ and $K$ band respectively, with their corresponding errors. The parameter range is chosen to be the same as from \cite{Mroz19}, with $m_{\rm s} < 21$ (in each band) and $\langle t_{\rm E} \rangle < 300$ days. The full range of $\langle \mu_{\rm rel} \rangle$ is used in those graphs. The increased effect of dust in the Galactic plane is evident at shorter wavelength, as well as the change in optical depth.
\newline\indent Later graphs show various cuts in $m_{\rm s}$, $\langle t_{\rm E} \rangle$ and $\langle \mu_{\rm rel} \rangle$ in $\tau$ and $\langle t_{\rm E} \rangle$. The effects of selecting different survey magnitude ranges on the optical depth and timescale are shown in figures \ref{figure:figure6} and \ref{figure:figure7}. The effect of reduced survey sensitivity (brighter $m_{\rm lim}$) is evident from the thicker dust bar; this is due to the dependence of source distance on apparent stellar brightness, as fewer lenses are likely to exist between the observer and a close star than for a more distant star. The effect on the timescale maps is also seen as a thinning of the bar as one selects fainter stars, however the lower bound of the timescale also drops as we select fainter stars. This is likely due to selecting stars with smaller Einstein radii, as well as those with a larger velocity dispersion (such as those in the bar), which result in a larger $\langle \mu_{\rm rel} \rangle$ relative to $\theta_{\rm E}$.

The effects of selecting different timescale ranges for optical depth and timescale maps are shown in figures \ref{figure:figure8} and \ref{figure:figure9}. Optical depth maps show a thickening of the dust bar for longer timescales and also a noticeable increase in noise due to the lower statistics. For timescale lower bounds of $\langle t_{\rm E} \rangle < 25$ days, the timescale maps respect the bounds chosen; however, as one picks timescale cuts above this range, the minimum timescale observed on the map can dip significantly below the bounds chosen. This is due to an artifact of the linear interpolation method, which struggles to preserve parameter bounds when interpolating between bins with significantly different statistics and weights.

Finally, the cuts in relative proper motion are shown in figures \ref{figure:figure10} and \ref{figure:figure11}. Choosing the lowest cuts in proper motion provided the highest optical depth values, which is consistent with the linear weighting of optical depth by timescale. The effect on the dust bar is subtle, as it is not as obviously thicker as the proper motion gets higher, although it does become more defined. The effect on timescale is prominent, as higher relative proper motion results naturally in a smaller timescale, which is evident from the range in values shown in figures. There is also evidence of the near side of the bar in the slowest proper motion cut, which is seen as a local increase in optical depth centred around $l=3^{\circ}$ with a width of $\sim 5^{\circ}$, which is consistent with the orientation and scale lengths of the bar outlined in section \ref{section:bar}. This asymmetry is mirrored by the average timescale, which is larger around the $l = 357^{\circ}$ region (shown as $l=-3^{\circ}$ in the figure). The primary contributor to this asymmetry is Bar-Disk lensing, which contributes 40\% of the total optical depth along this line of sight.

There is also evidence of a cross-like structure in the highest relative proper motion cut of the timescale maps in figure \ref{figure:figure11}. This is likely caused by two competing factors. In the horizontal direction, outside the influence of the bulge, timescale tends to increase away from the Galactic centre in longitude as the transverse component of the disk rotational velocity projected onto our line of sight is smaller, resulting in a longer timescale. The timescale in Galactic latitude also increases away from the Galactic centre, as source stars tend to be closer to the observer due to the scale height of the disk. Due to this closer distance, $\theta_E$ tends to be larger, also increasing the timescale. The relative proper motion cut changes how these two factors contribute to the overall structure of the timescale map.

\subsection{The brown dwarf mass function}

\begin{figure}
    \centering
    \includegraphics[width=7.5cm]{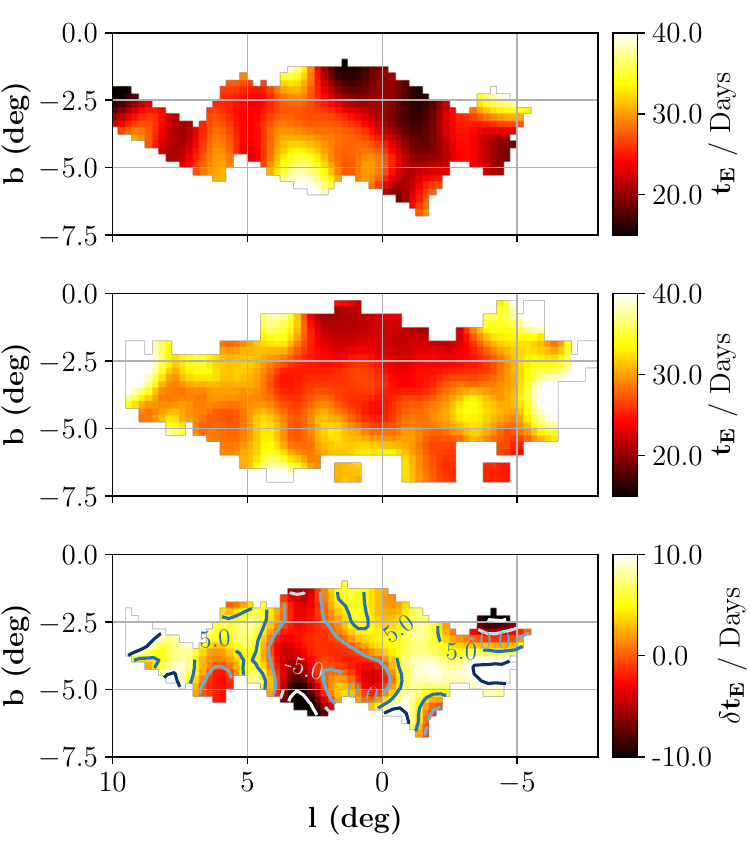}
    \caption{Maps of $\langle t_{\rm E} \rangle$ for the MOA-II survey (top, \citet{Sumi13}), OGLE-IV survey (middle, \citet{Mroz19}) and the difference between them (bottom, OGLE - MOA). The mean difference in timescale is an excess of 3.16 days for the OGLE-IV data. For comparisons' sake, the positive latitude portion of the OGLE-IV timescale map is not shown, as the MOA-II map only includes data in the negative latitude sky. The resolution of both maps was fixed at $0.25^\circ$.}
    \label{fig:timescale}
\end{figure}

\begin{figure*}
    \centering
    \includegraphics[width=.9\textwidth,height=.9\textheight]{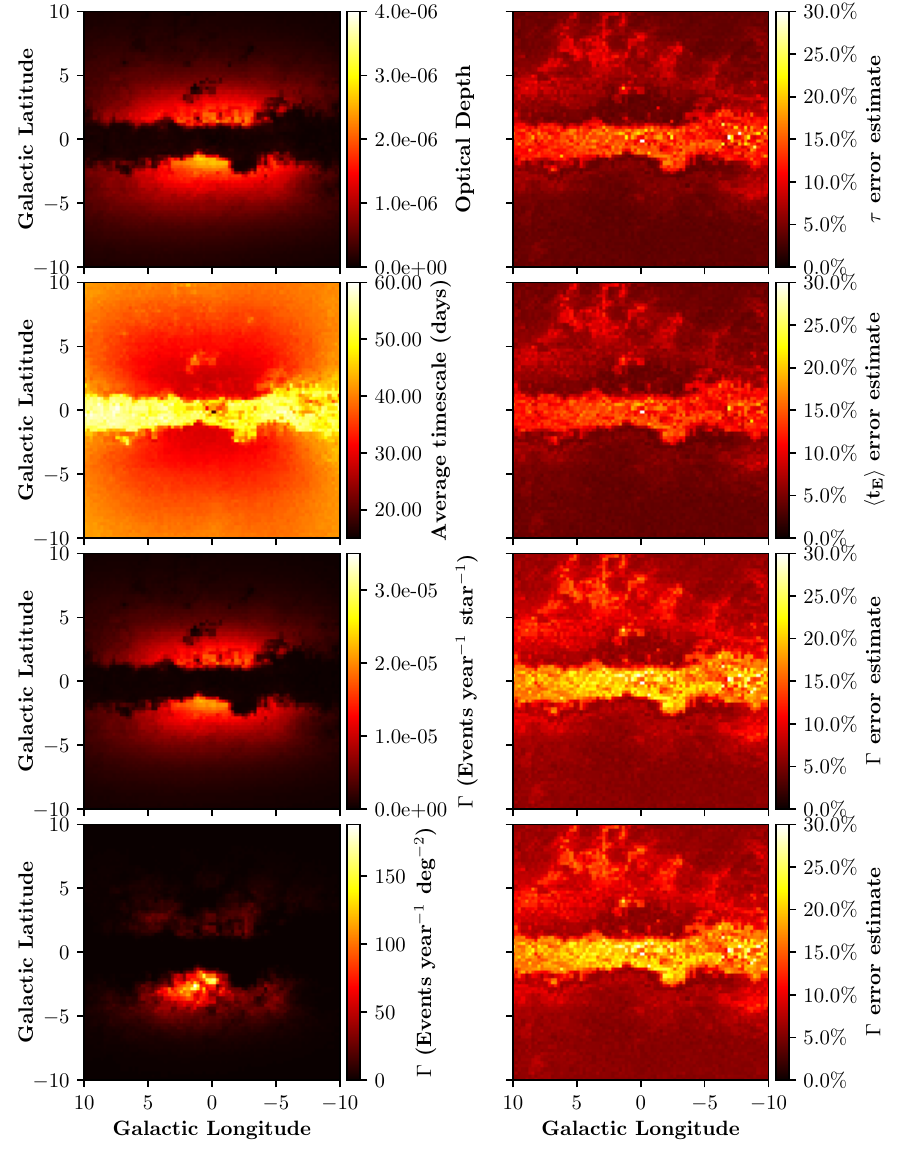}
    \caption{$V$-band parameter maps are shown on the left column, with associated percent errors on the right column. Parameter ranges for all plots are $V < 21$, $\langle t_{\rm E} \rangle < 300$ days and $\langle \mu_{\rm rel} \rangle < 20$ mas year$^{-1}$. From top to bottom, the parameter maps are: the microlensing optical depth $\tau$, the average Einstein radius crossing time $\langle t_{\rm E} \rangle$, the event rate per source star and the event rate per square degree, $\Gamma$.}
    \label{figure:figure3}
\end{figure*}

\begin{figure*}
    \centering
    \includegraphics[width=.9\textwidth,height=.9\textheight]{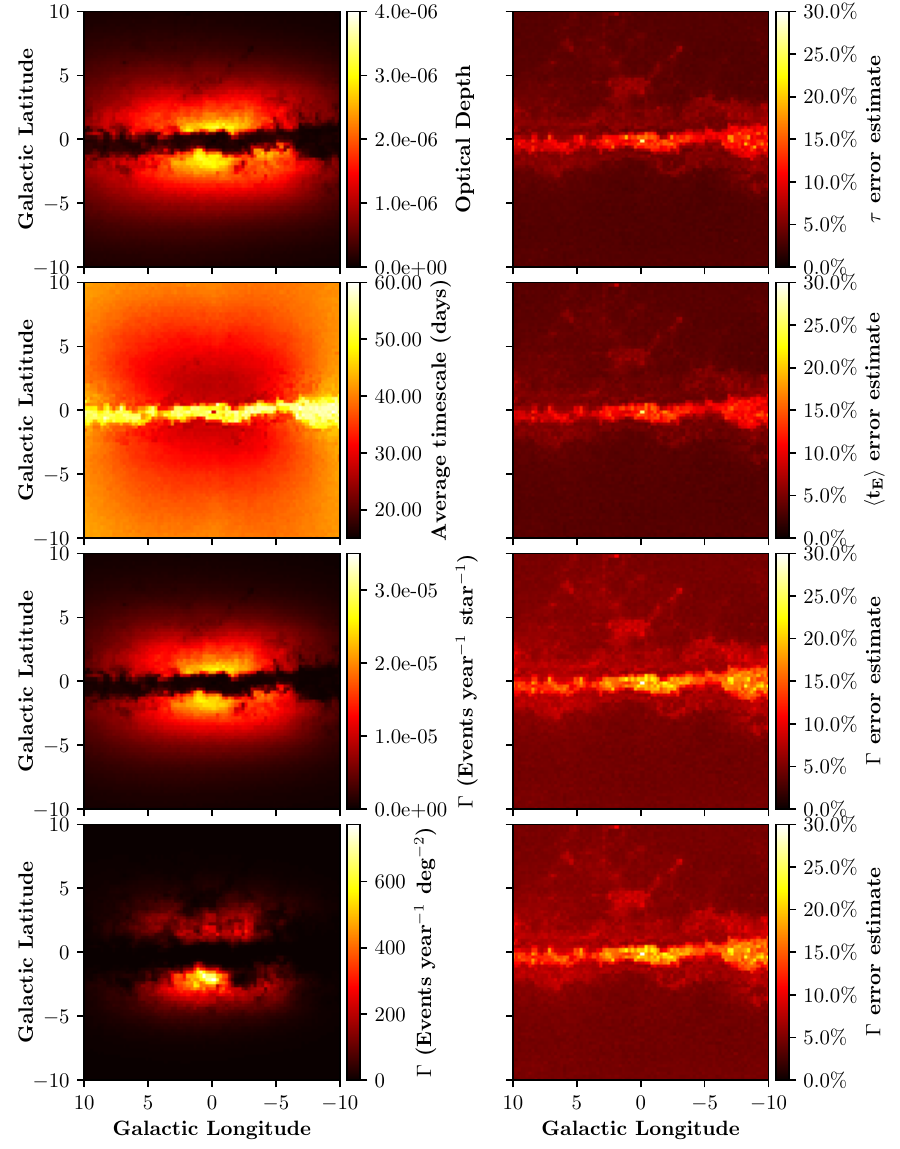}
    \caption{$I$-band parameter maps are shown on the left column, with associated percent errors on the right column. Parameter ranges for all plots are $I < 21$, $\langle t_{\rm E} \rangle < 300$ days and $\langle \mu_{\rm rel} \rangle < 20$ mas year$^{-1}$. From top to bottom, the parameter maps are: the microlensing optical depth $\tau$, the average Einstein radius crossing time $\langle t_{\rm E} \rangle$, the event rate per source star and the event rate per square degree, $\Gamma$.}
    \label{figure:figure4}
\end{figure*}

\begin{figure*}
    \centering
    \includegraphics[width=.9\textwidth,height=.9\textheight]{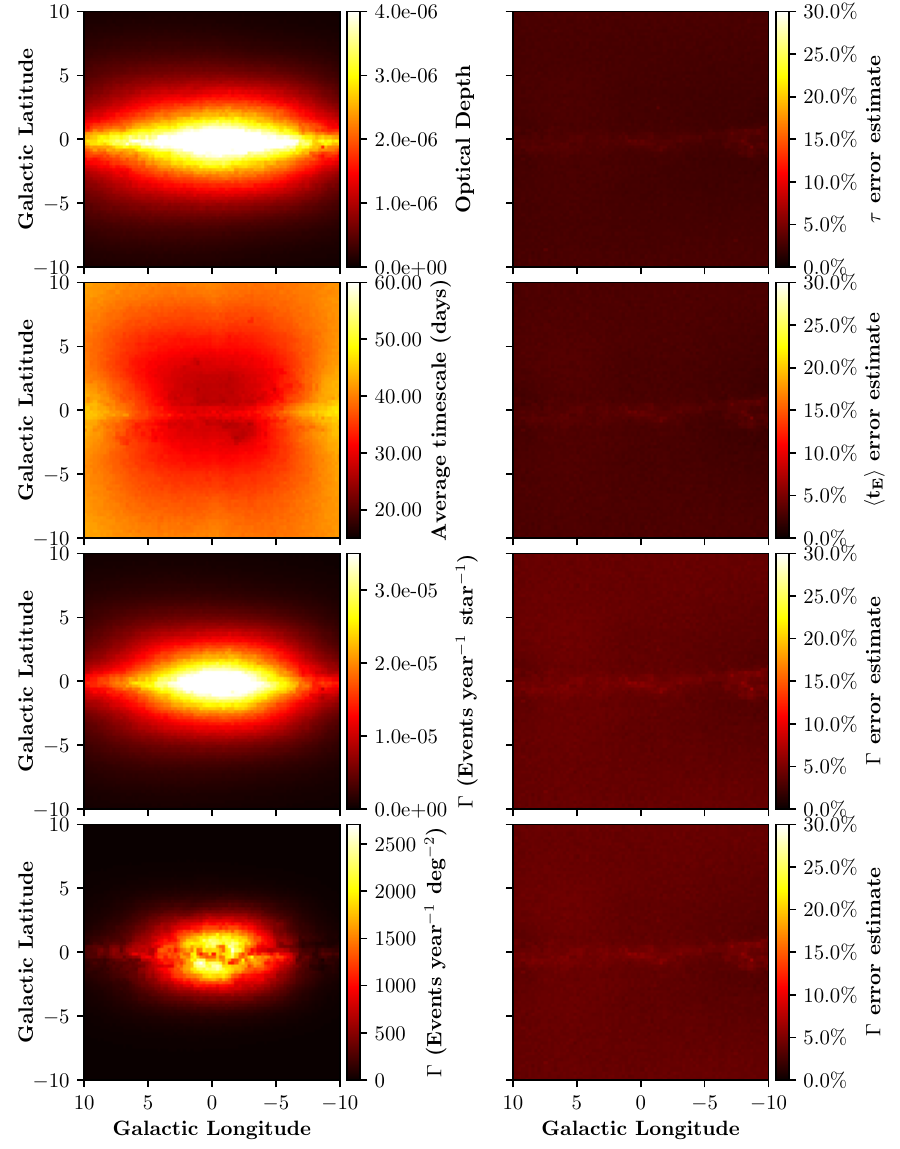}
    \caption{$K$-band parameter maps are shown on the left column, with associated percent errors on the right column. Parameter ranges for all plots are $K < 21$, $\langle t_{\rm E} \rangle < 300$ days and $\langle \mu_{\rm rel} \rangle < 20$ mas year$^{-1}$. From top to bottom, the parameter maps are: the microlensing optical depth $\tau$, the average Einstein radius crossing time $\langle t_{\rm E} \rangle$, the event rate per source star and the event rate per square degree, $\Gamma$.}
    \label{figure:figure5}
\end{figure*}

\begin{figure*}
    \centering
    \includegraphics[width=.9\textwidth,height=.9\textheight]{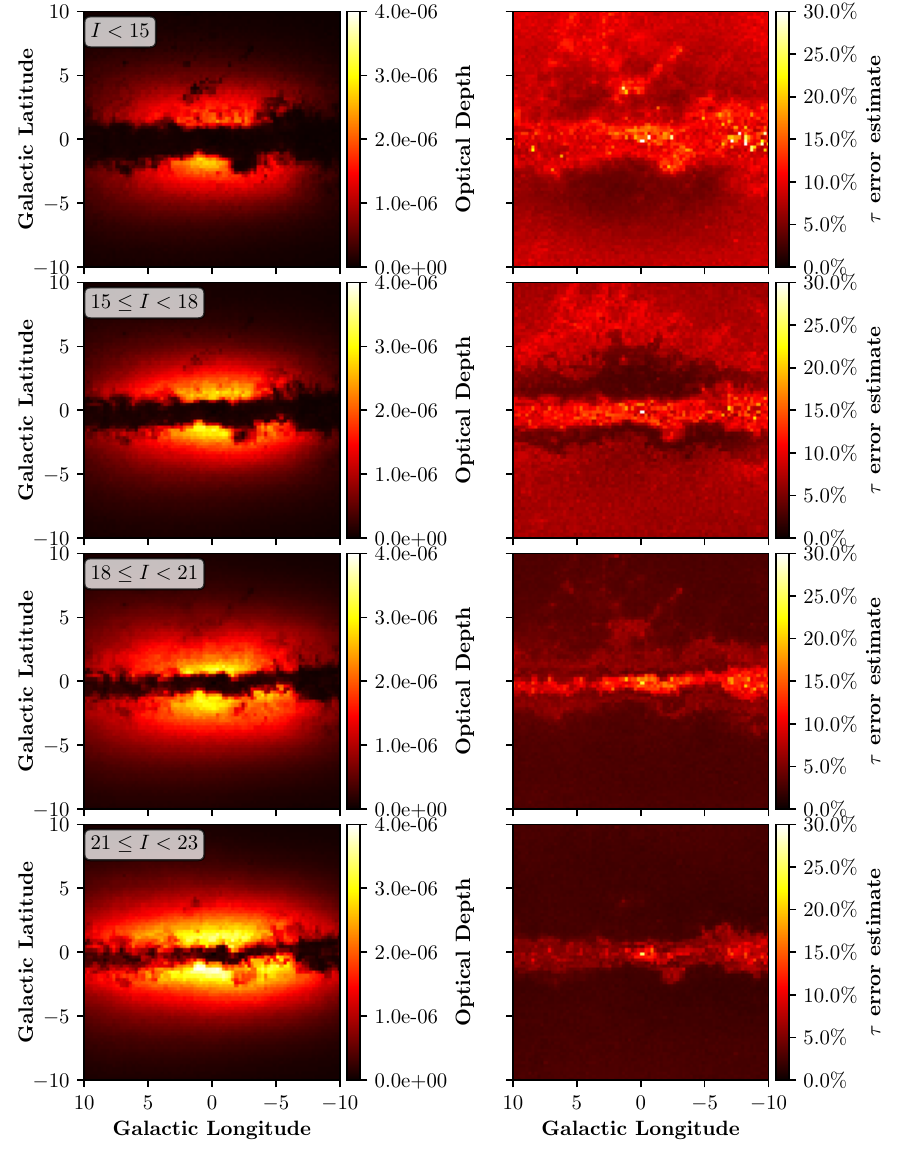}
    \caption{The optical depth in $I$-band is shown over various survey magnitude ranges with associated error. From top to bottom, these ranges are $I < 15$, $15 \leq I < 18$, $18 \leq I < 21$ and $21 \leq I < 23$. The timescale and relative proper motion ranges are kept as $\langle t_{\rm E} \rangle < 300$ days and $\langle \mu_{\rm rel} \rangle < 20$ mas year$^{-1}$. The thinning dust bar and rising optical depth is evident as the magnitude cut selects fainter stars.}
    \label{figure:figure6}
\end{figure*}

\begin{figure*}
    \centering
    \includegraphics[width=.9\textwidth,height=.9\textheight]{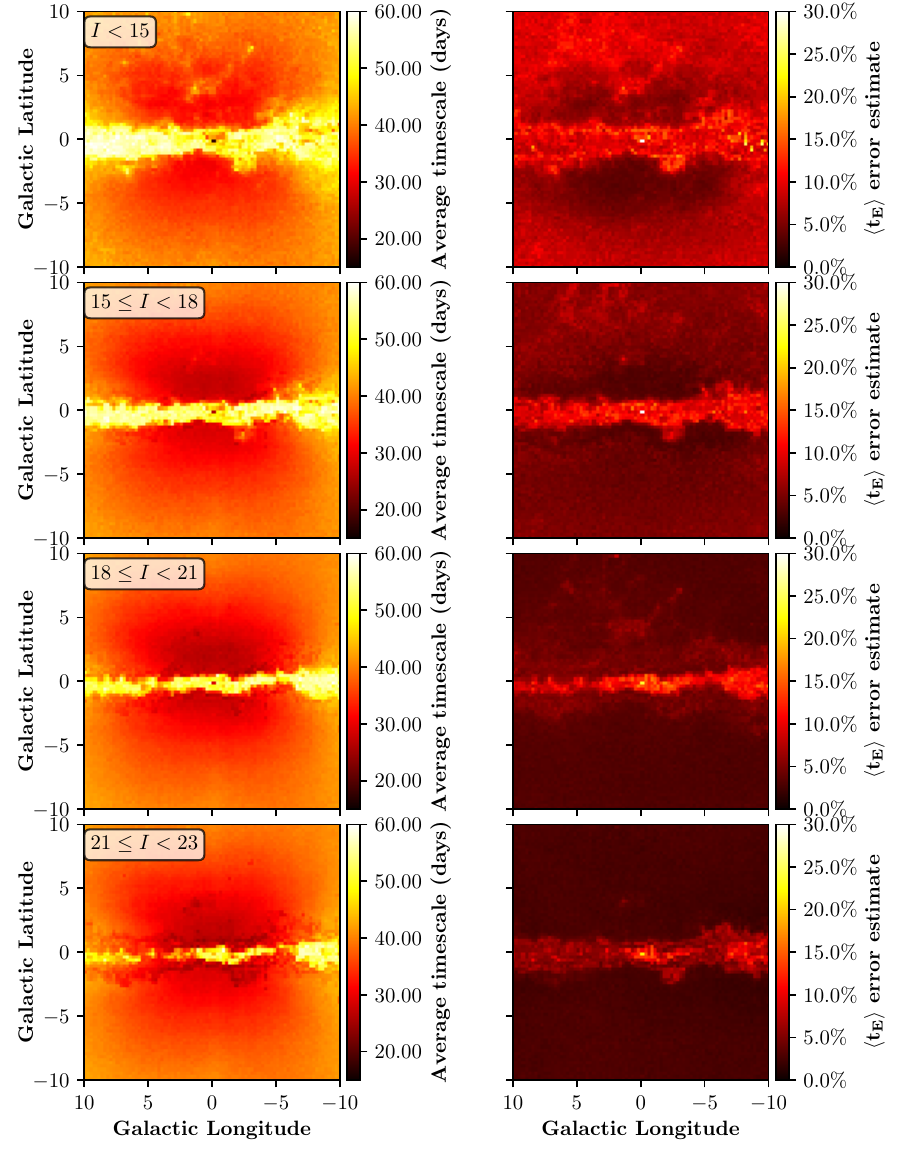}
    \caption{The average timescale in $I$-band is shown over various survey magnitude ranges with associated error. From top to bottom, these ranges are $I < 15$, $15 \leq I < 18$, $18 \leq I < 21$ and $21 \leq I < 23$. The timescale and relative proper motion ranges are kept as $\langle t_{\rm E} \rangle < 300$ days and $\langle \mu_{\rm rel} \rangle < 20$ mas year$^{-1}$. The falling lower timescale bound is evident as the magnitude cut selects fainter stars.}
    \label{figure:figure7}
\end{figure*}

\begin{figure*}
    \centering
    \includegraphics[width=.9\textwidth,height=.9\textheight]{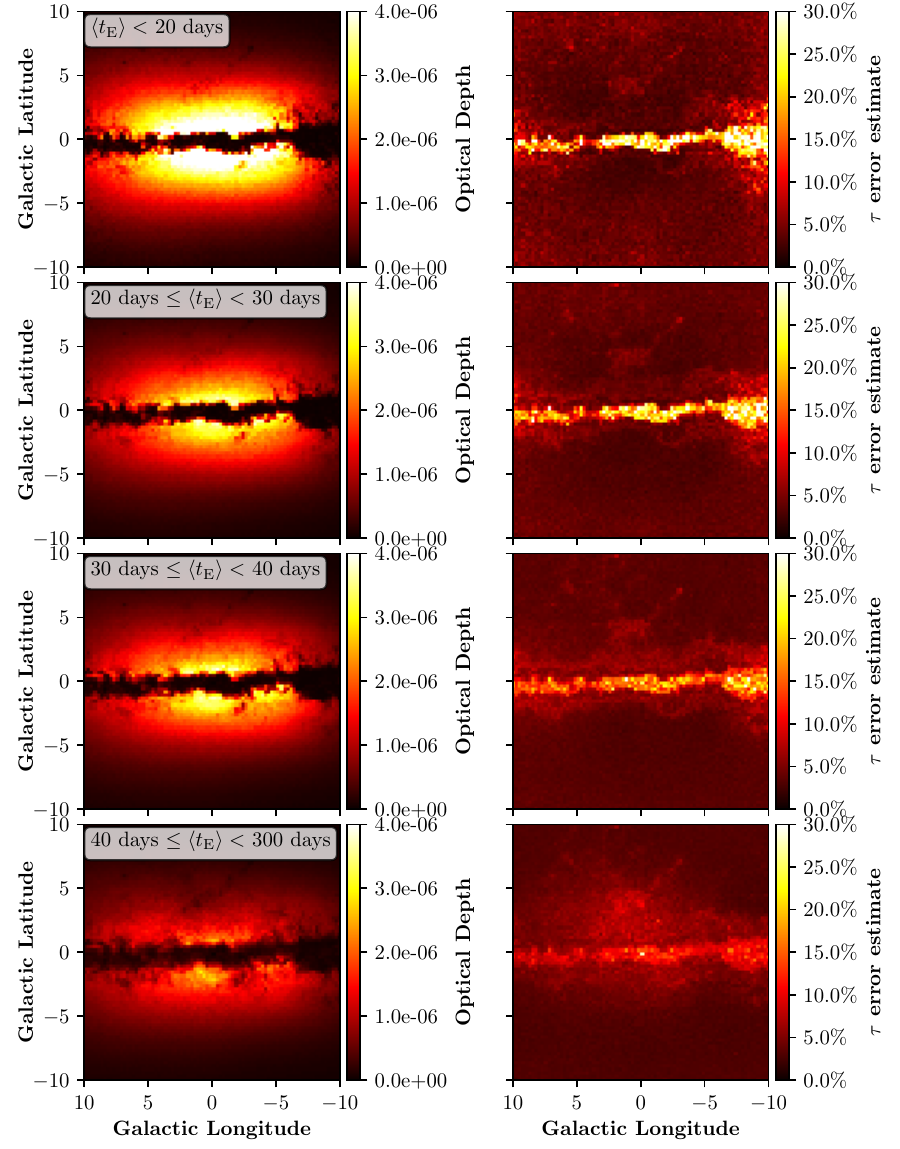}
    \caption{The optical depth in $I$-band is shown over various timescale ranges with associated error. From top to bottom, these ranges are $\langle t_{\rm E} \rangle < 20$ days, 20 days $\leq \langle t_{\rm E} \rangle < 30$ days, 30 days $\leq \langle t_{\rm E} \rangle < 40$ days and 40 days $\leq \langle t_{\rm E} \rangle < 300$ days. The magnitude and relative proper motion ranges were kept constant at $I < 21$ and $\langle \mu_{\rm rel} \rangle < 20$ mas year$^{-1}$.}
    \label{figure:figure8}
\end{figure*}

\begin{figure*}
    \centering
    \includegraphics[width=.9\textwidth,height=.9\textheight]{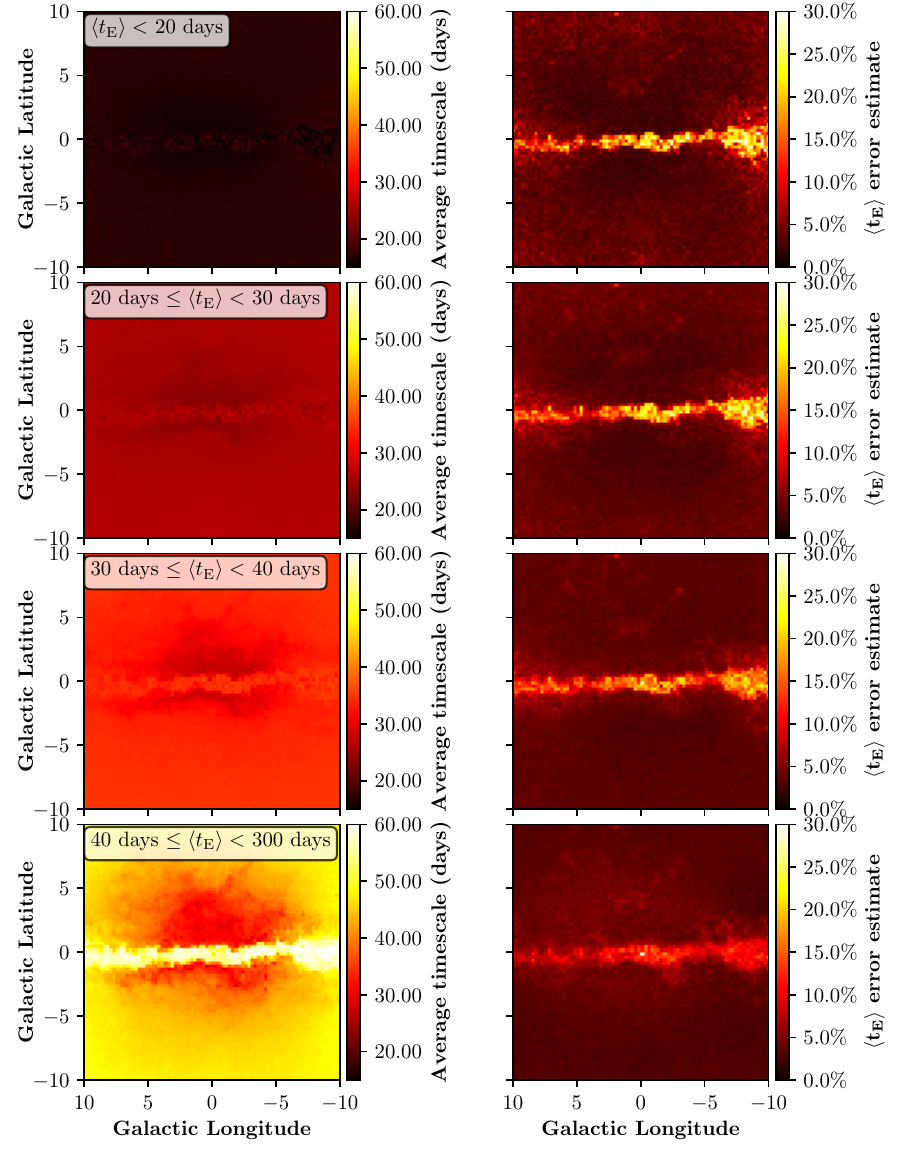}
    \caption{The average timescale in $I$-band is shown over various timescale ranges with associated error. From top to bottom, these ranges are $\langle t_{\rm E} \rangle < 20$ days, 20 days $\leq \langle t_{\rm E} \rangle < 30$ days, 30 days $\leq \langle t_{\rm E} \rangle < 40$ days and 40 days $\leq \langle t_{\rm E} \rangle < 300$ days. The magnitude and relative proper motion ranges were kept constant at $I < 21$ and $\langle \mu_{\rm rel} \rangle < 20$ mas year$^{-1}$. The interpolation method struggles to preserve the chosen timescale bounds as the range increases, as evident in the lower two rows. }
    \label{figure:figure9}
\end{figure*}

\begin{figure*}
    \centering
    \includegraphics[width=.9\textwidth,height=.9\textheight]{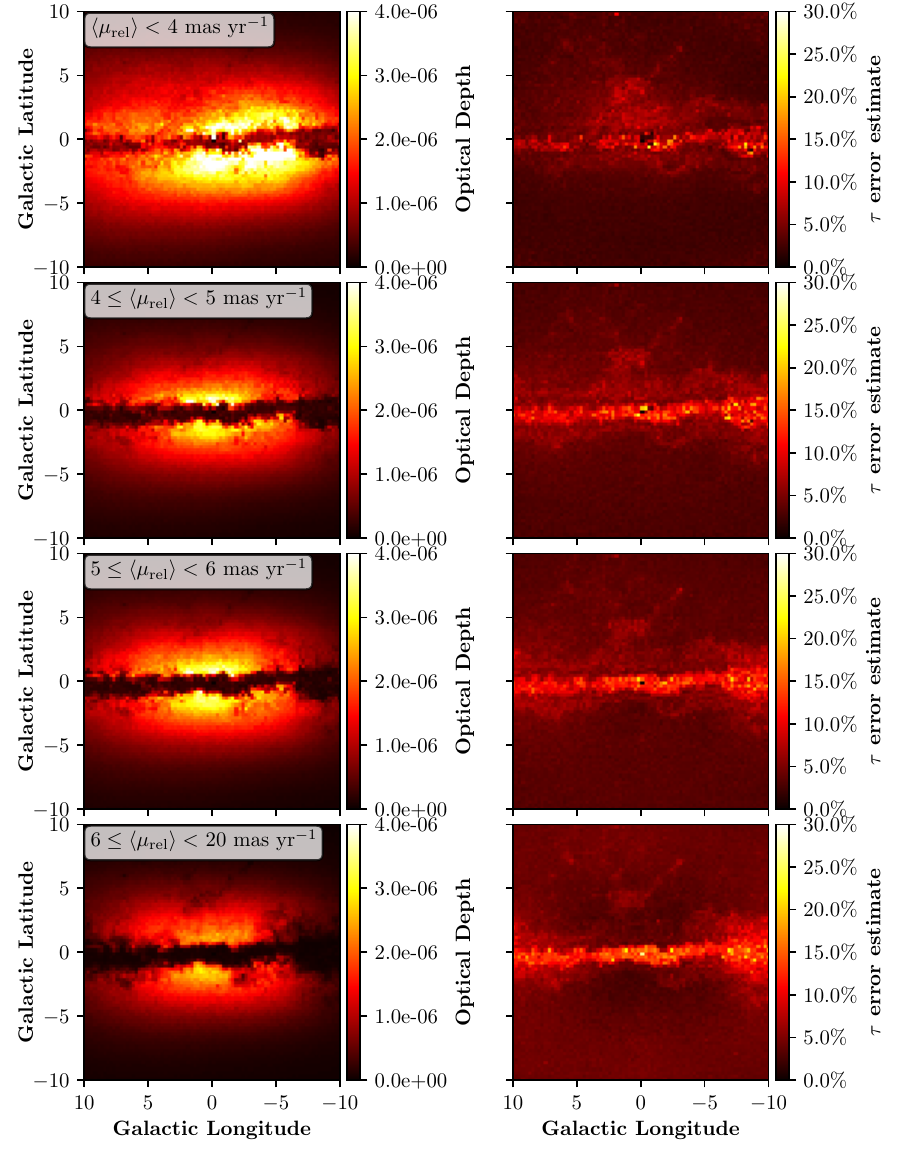}
    \caption{The optical depth in $I$-band is shown over various relative proper motion ranges with associated error. From top to bottom, these ranges are $\langle \mu_{\rm rel} \rangle < 4$ mas year$^{-1}$, 4 mas year$^{-1} \leq \langle \mu_{\rm rel} \rangle < 5$ mas year$^{-1}$, 5 mas year$^{-1} \leq \langle \mu_{\rm rel} \rangle < 6$ mas year$^{-1}$, 6 mas year$^{-1} \leq \langle \mu_{\rm rel} \rangle < 20$ mas year$^{-1}$. The magnitude and average timescale ranges were kept constant at $I < 21$ and $\langle t_{\rm E} \rangle < 300$ days. The nearside of the bar is evident in the lowest proper motion cut, where the optical depth is larger around $l=-3^{\circ}$.}
    \label{figure:figure10}
\end{figure*}

\begin{figure*}
    \centering
    \includegraphics[width=.9\textwidth,height=.9\textheight]{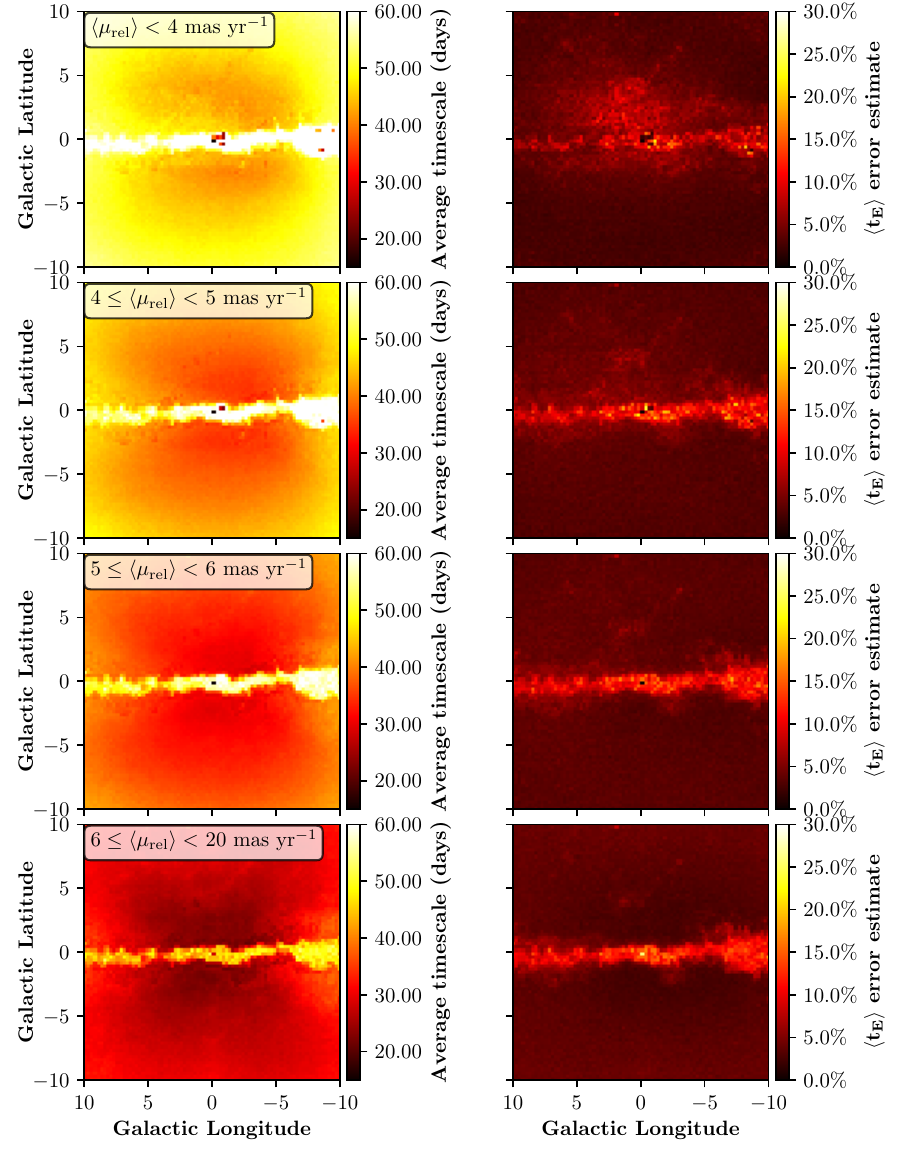}
    \caption{The average timescale in $I$-band is shown over various relative proper motion ranges with associated error. From top to bottom, these ranges are $\langle \mu_{\rm rel} \rangle < 4$ mas year$^{-1}$, 4 mas year$^{-1} \leq \langle \mu_{\rm rel} \rangle < 5$ mas year$^{-1}$, 5 mas year$^{-1} \leq \langle \mu_{\rm rel} \rangle < 6$ mas year$^{-1}$, 6 mas year$^{-1} \leq \langle \mu_{\rm rel} \rangle < 20$ mas year$^{-1}$. The magnitude and average timescale ranges were kept constant at $I < 21$ and $\langle t_{\rm E} \rangle < 300$ days.}
    \label{figure:figure11}
\end{figure*}

The discrepancy of the fitted brown dwarf mass function slopes between the previous work by \cite{Awiphan16} and this work is due mainly to the difference in the recovered timescale distributions for the lowest timescale regime between the MOA-II and OGLE-IV surveys. This has been noted in the context of FFP analyses in \citet{Sumi11} and \citet{Mroz17}. Figure \ref{fig:timescale} shows this discrepancy, with a noticeable excess in the OGLE-IV timescale map in most parts other than around $2 < l < 4$, with an average excess of 3.16 days across the overlapping region (shown in the bottom of figure \ref{fig:timescale}). As the mass function slope was fitted by sampling across the entire OGLE-IV timescale map, fewer simulated brown dwarfs were ultimately required to bring the timescale in accordance with the OGLE-IV results.

\begin{figure*}
    \centering
    \includegraphics[width=9cm]{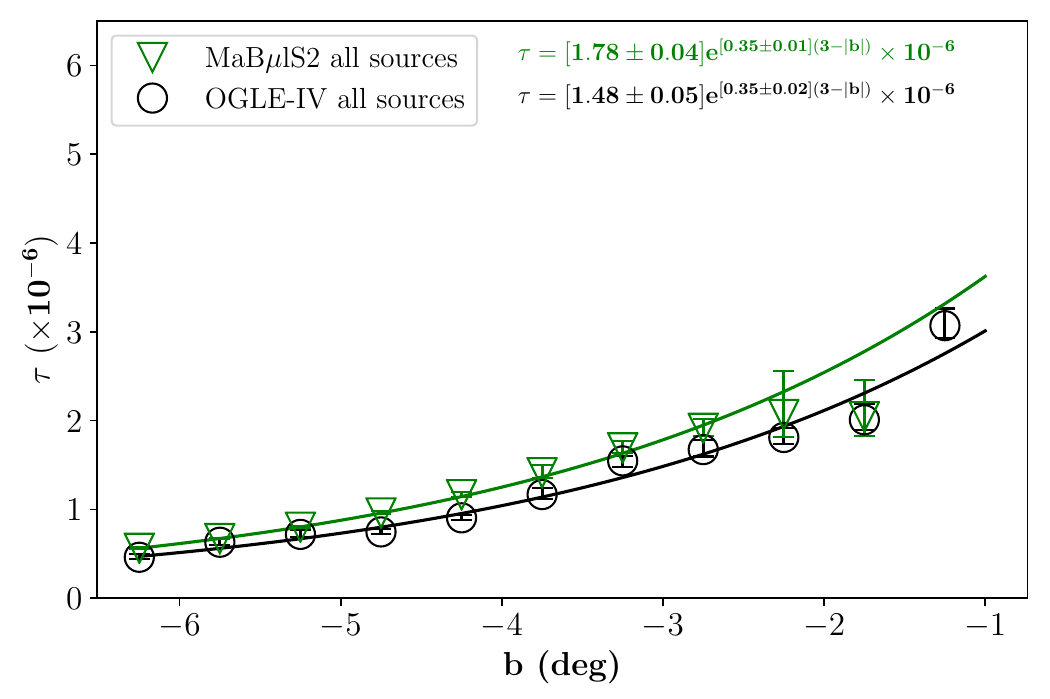}
    \caption{The comparison between the optical depth variations of OGLE-IV source stars from \citet{Mroz19} over the longitude range $|l| < 5^\circ$ as well as the new \mabuls{}2 result from the simulation. In black is the OGLE-IV data, binned in intervals of $0.5^\circ$ in galactic latitude. \mabuls{}-2 is shown in green. The lines show the optimal exponential fits for their respective colours, with equations shown at the top of the figure.}
    \label{fig:ogle_compare}
\end{figure*}

\begin{figure*}
    \centering
    \includegraphics[width=9cm]{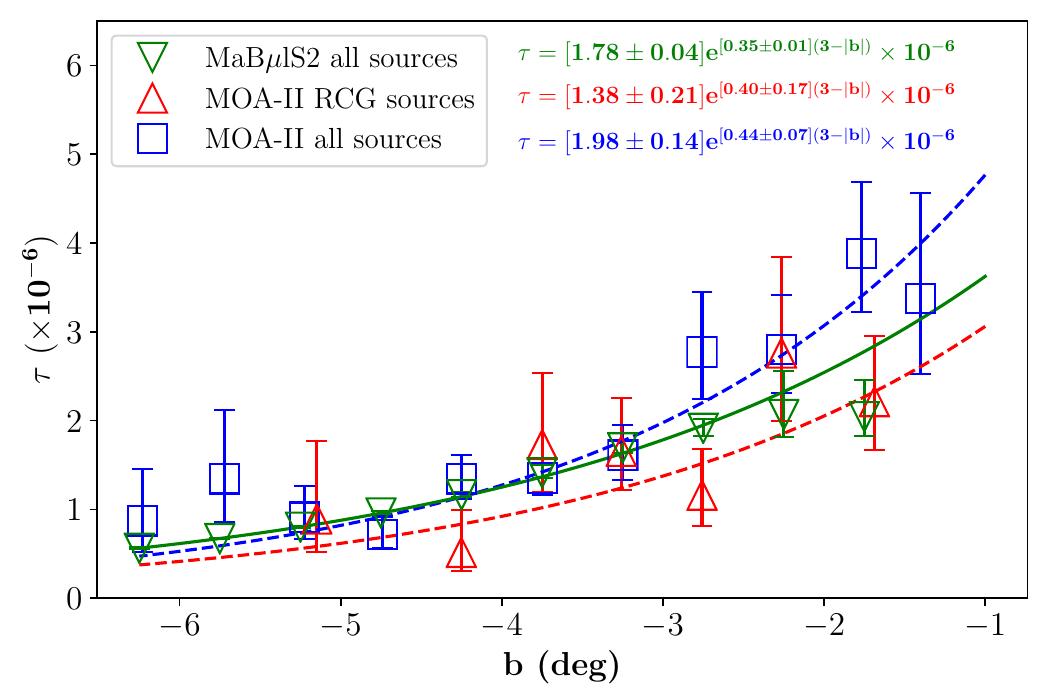}
    \caption{The comparison between the optical depth variations of MOA-II source stars (all sources and exclusively RCG sources) from \citet{Sumi2016} over the longitude range $|l| < 5^\circ$ as well as the new \mabuls{}2 result from the simulation. The MOA-II results have been scaled by a factor $1/0.931$ to account for the differing telescope conditions between OGLE-IV and MOA-II, mentioned in section \ref{section:bg_light}. In red is the MOA-II RCG data, with the fit for all sources shown in blue, each binned in intervals of $0.5^\circ$ in galactic latitude. \mabuls{}-2 is shown in green. The lines show the optimal exponential fits for their respective colours, with equations shown at the top of the figure.}
    \label{fig:moa_compare}
\end{figure*}

There is good agreement between the \mabuls{} optical depth profile and the OGLE-IV survey near the galactic bulge, as evident in figure \ref{fig:ogle_compare}. Comparing the optical depth between MOA-II and OGLE-IV shows more similarity between the resolved RCG exponential than the all-sources exponential from MOA-II results \citep{Sumi2016}, as shown in figure \ref{fig:moa_compare}. This suggests that strong blending could be the source of tension between the two brown dwarf slopes.

\subsection{Comparison with OGLE IV data}\label{section:ogledata}

\begin{figure*}
    \centering
    \includegraphics[width=.9\textwidth,height=.9\textheight]{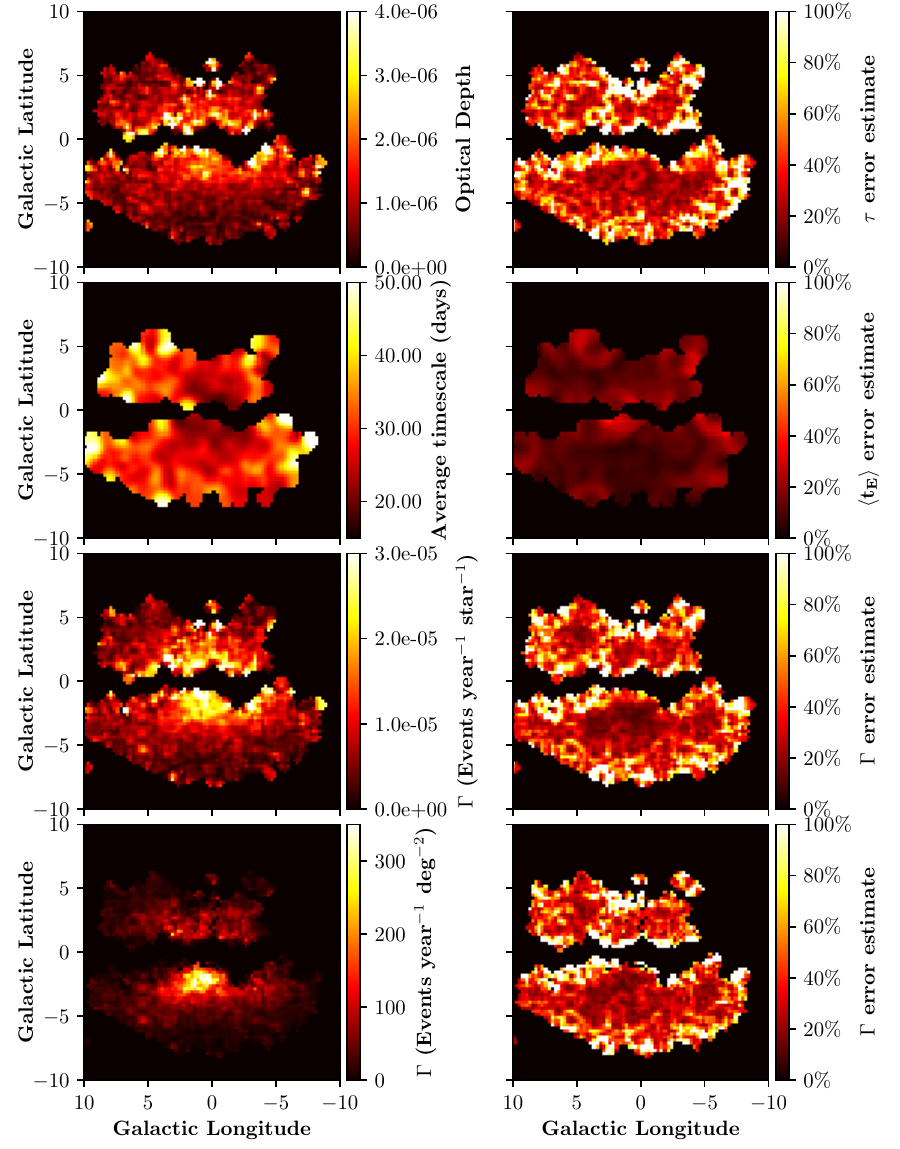}
    \caption{Data compiled from the OGLE IV analysis between the years 2010 to 2017. The parameter ranges used are $14 < I < 21$, $\langle t_{\rm E} \rangle < 300$ days. The corresponding error maps shown are the errors on the residual between the OGLE survey and the \mabuls{}-2 model, which for a perfect model would be an accurate representation of the error. Given that the model is imperfect, these errors are approximate.}
    \label{figure:figure12}
\end{figure*}

\begin{figure*}
    \centering
    \includegraphics[width=.9\textwidth,height=.9\textheight]{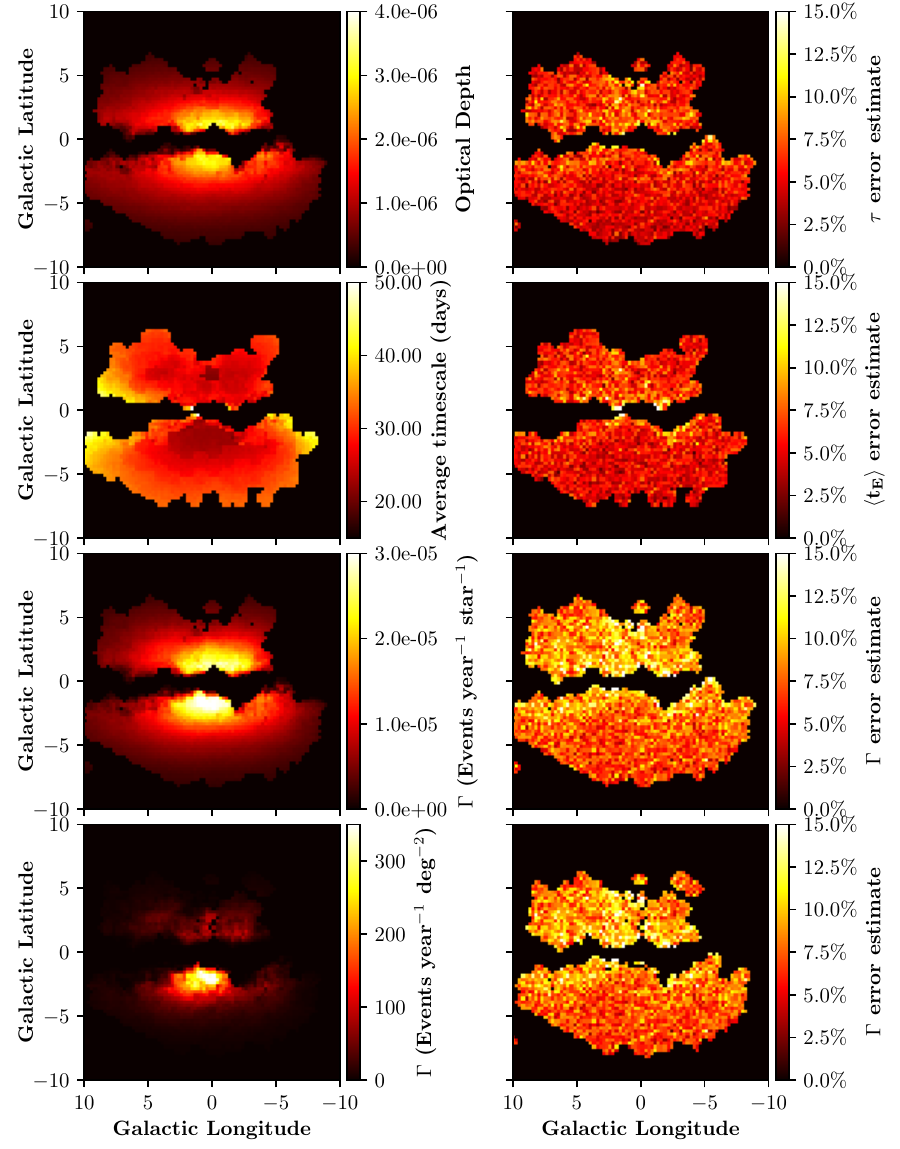}
    \caption{A \mabuls{}-2 simulation using the same event selection criteria outlined in \citet{Mroz19}, with a mask applied over the regions not applicable in the OGLE data. The parameter ranges used are $14 < I < 21$, $\langle t_{\rm E} \rangle < 300$ days. The signal-to-noise selection criterion characterised by equations \ref{equation:equation22} and \ref{equation:equation23} was modified to match the time-integrated signal to noise from equation \ref{equation:equation25}. The sharp contours in the timescale map are due to the differing OGLE field cadences, and the effect this has on the signal to noise criterion.}
    \label{figure:figure13}
\end{figure*}

\begin{figure*}
    \centering
    \includegraphics[width=.9\textwidth,height=.9\textheight]{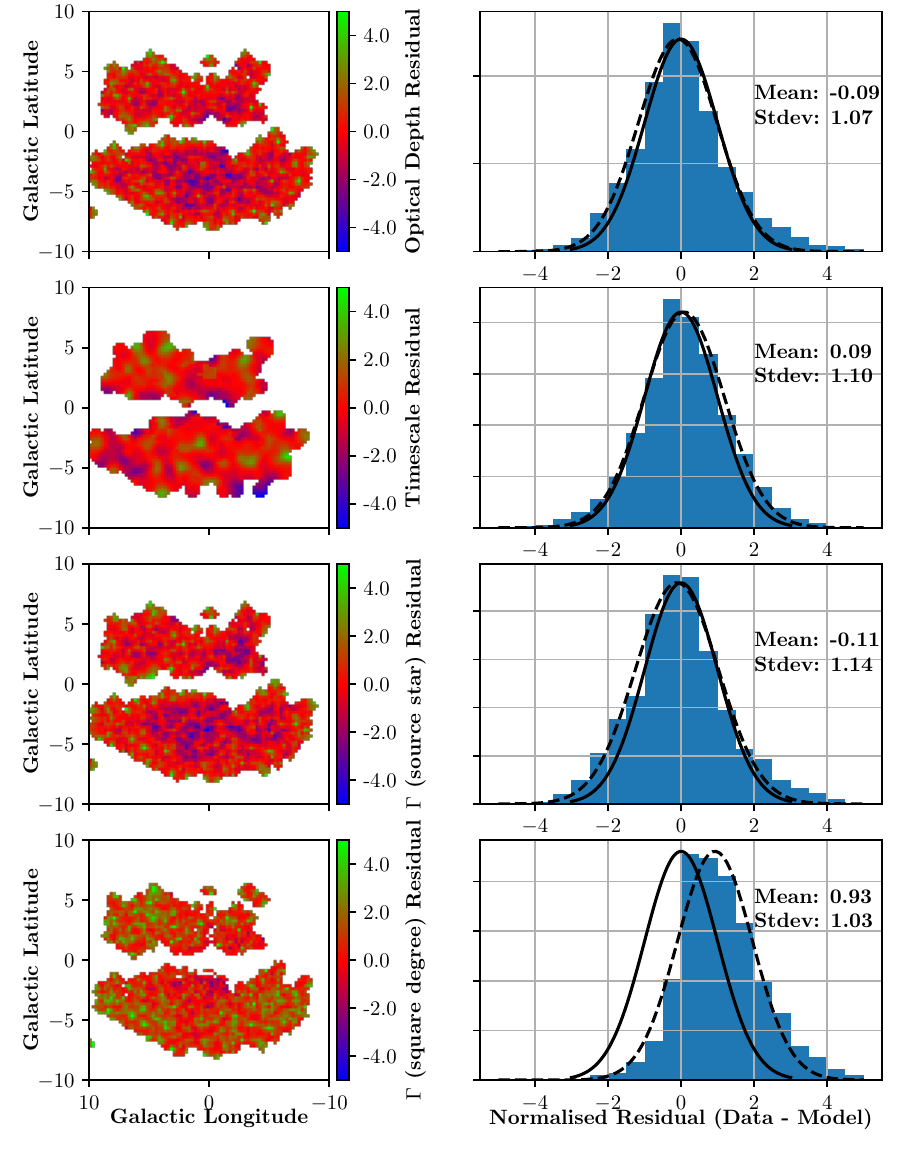}
    \caption{The residuals (OGLE - Model) for each of the parameter maps shown in figure \ref{figure:figure12}. Residuals have been normalised to their variance. The left column shows the map of residuals, with the corresponding 1D histograms in the right column. Two Gaussians are plotted on each of the histograms. The solid black Gaussian is a unit Gaussian with $\mu = 0$ and $\sigma = 1$, representing the ideal scenario where the model matches the data. The dashed Gaussian is fitted to the distribution with the mean and standard deviation listed on the plot.}
    \label{figure:figure14}
\end{figure*}

\begin{figure*}
    \centering
    \includegraphics[width=.9\textwidth,height=.9\textheight]{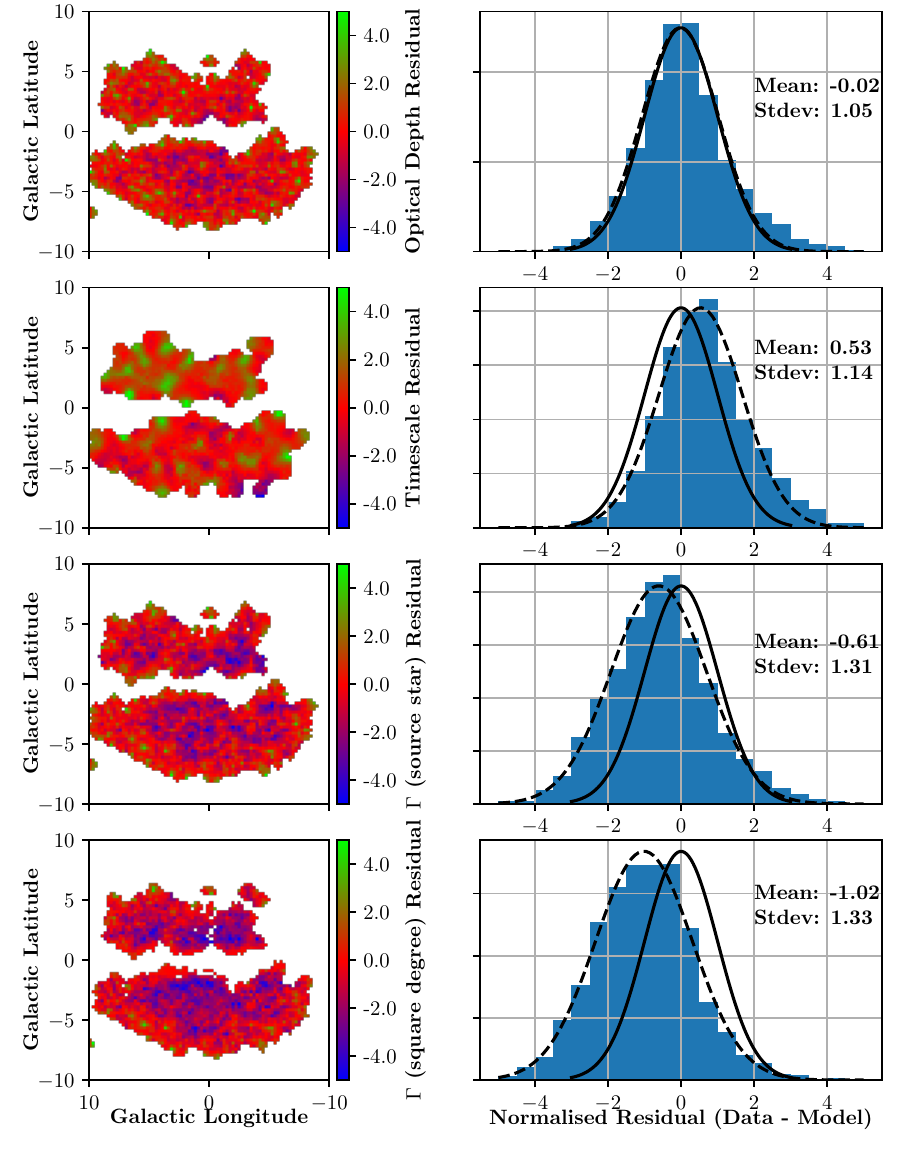}
    \caption{The residuals (OGLE - Model) for each of the parameter maps shown in figure \ref{figure:figure12}, for the old version of \mabuls{} \citep{Awiphan16}. Residuals have been normalised to their variance. The left column shows the map of residuals, with the corresponding 1D histograms in the right column.}
    \label{figure:figure15}
\end{figure*}

The OGLE-IV parameter maps are shown in figure \ref{figure:figure12} with their associated error maps\footnote{The OGLE-IV parameter map and microlensing event data is available at \url{http://ogle.astrouw.edu.pl/cgi-ogle/get_o4_tau.py}}. A modified version of \mabuls{} was run which more closely matched the event selection criteria of the OGLE-IV survey, including an accurate treatment of the field cadences, with results shown for $I$-band in figure \ref{figure:figure13}. The signal-to-noise criterion for OGLE-IV outlined in \cite{Mroz19} was a time integrated one, as opposed to requiring a particular signal-to-noise threshold at peak magnification (equation \ref{equation:equation22}). It required that the sum over all the flux difference at least $3\sigma$ above baseline, normalised to the scatter, was at least 32,
\begin{equation}
    \label{equation:equation25}
    \chi_{3+} = \sum_{i}\frac{F_i-F_{\rm base}}{\sigma_i} \geq 32
\end{equation}
The residuals (OGLE - model), normalised to their errors are shown in figure \ref{figure:figure14}, with equivalent residuals shown for the older version of \mabuls{} (brown dwarfs included) shown in figure \ref{figure:figure15}. The error on the residual was calculated by finding the standard deviation of all nearby points within a donut shaped kernel, with an inner radius $r_{\rm inner} = 15'$ for optical depth, rate per source star and rate per square degree and $r_{\rm inner} = 30'$ for the average timescale and an outer radius $r_{\rm outer} = 3r_{\rm inner}$. The inner radius prevents correlating the error estimate with the parameter value, while the outer radius prevents sampling points too far away from the target location to act as representative of the local region. The resulting distribution of residuals is an estimate of the accuracy of the model; in the ideal scenario, the distributions would be unit gaussians with a mean of zero. Offsets in the mean and standard deviation thus suggest inaccuracies in the model.

The residual maps suggest that the model is consistent with the data, with some spatial variations visible in the optical depth map, with a notable over-prediction in the bar. This discrepancy is propagated to the rate per source star. The over-prediction in $\tau$ is similar to the original \mabuls{} residual map, which shows similar features. The new timescale residual map suggests strong agreement with the data, with a marked improvement over the original, which under-predicted the timescale above the Galactic plane, although this is to be expected given that the previous model was optimised for the smaller MOA II sample with different event selection. The microlensing rate per source star shows good agreement in the residual histogram, with spatial variation in the residual map consistent with the propagation of the optical depth. This is opposed to the old \mabuls{}, which shows a much stronger over-prediction of the rate per source star, with a non unit Gaussian distribution of residuals. In both models, the rate per square degree is over predicted near the bar, although this effect is much more dominant in the previous version, which over-predicts the rate across much of the field. The new model shows a uniform under-prediction of the rate per square degree outside the bar, where the event rate drops below 100 events per year. The inconsistencies in the data with the rate per square degree could likely be due to inaccuracies with the IMFs used, which in turn influence the luminosity of the sources; changing the IMFs would not necessarily have a large effect on the optical depth or timescale, which depend on lens mass as $\sqrt{M}$, but luminosities would, with a stronger dependency on the order of $L(M) \propto M^\alpha$ with values of $\alpha$ in the range [2,4]. Alternatively, the discrepancy could also be caused by an insufficiently sophisticated source weighting which does not replicate the OGLE IV survey conditions.

\section{Conclusion}\label{section:conclusion}

A new microlensing model has been presented which develops substantially upon previous work. The inclusion of a formal finite source treatment, improved error calculations, background light contributions and lens-source relative proper motion cuts allows for improved microlensing optical depth, timescale and rate estimations. Calculating the distribution of normalised residuals for each microlensing parameter (optical depth, timescale, rate per source star and rate per square degree) showed improvements in all parameters over the old model, except for optical depth which remained the same. A notable discrepancy between the new model and the OGLE-IV results was seen in the rate per square degree, with a notable under prediction by the model outside of the bulge region. This discrepancy was the inverse of the old model, which displays a strong over prediction across the whole OGLE-IV rate map. Two possible solutions to the discrepancy are proposed, namely that the stellar IMFs (and by extension, luminosity functions) are inaccurate, leading to a smaller population of resolved stars, or that the source weighting used by the simulation is insufficiently faithful to the OGLE-IV survey, resulting in incorrect contributions from each source star. The brown dwarf mass function was fitted to the OGLE-IV data with a $\chi^2$ minimisation and the resulting mass function slope was found to be +0.1, contrasting significantly with previous work, suggesting a slope of -0.4, which was determined to be a result of the effect of differing event selection criteria between the OGLE-IV and MOA-II surveys on the resulting mean event timescale; as the OGLE-IV timescale map was on average higher than the MOA-II map, the necessity for short timescale brown dwarfs in the new model was lessened. \mabuls{}2 has the potential to be extended to incorporate space-based surveys such as Roman (formerly WFIRST) and Euclid as well as upcoming ground based surveys such as the Rubin Observatory (formerly the LSST).

\section*{Acknowledgements}

David Specht is funded by a UK Science and Technology Facilities Council (STFC) PhD studentship. Eamonn Kerins also acknowledges funding from STFC. We would also like to thank the anonymous referee for their input, much of which has been incorporated into this paper.

\section*{Data Availability}

\mabuls-2 is available online at \url{www.mabuls.net} to provide on-the-fly maps for user supplied cuts in survey magnitude, event timescale and relative proper motion.

%%%%%%%%%%%%%%%%%%%%%%%%%%%%%%%%%%%%%%%%%%%%%%%%%%



\bibliographystyle{mnras} 
\bibliography{ref}

%%%%%%%%%%%%%%%%%%%%%%%%%%%%%%%%%%%%%%%%%%%%%%%%%%

% Don't change these lines
\bsp	% typesetting comment
\label{lastpage}
\end{document}